\documentclass[sigconf]{acmart}

\renewcommand\footnotetextcopyrightpermission[1]{} 
\AtBeginDocument{%
  \providecommand\BibTeX{{%
    \normalfont B\kern-0.5em{\scshape i\kern-0.25em b}\kern-0.8em\TeX}}}

\usepackage{enumitem}

\usepackage{kg-macros}
\usepackage{mathtools}
\usepackage{listings} 

\usepackage{todonotes}

\usepackage{pgfplots}

\usepackage{booktabs}

\newcommand{\data}{domain graph\xspace}
\newcommand{\datas}{domain graphs\xspace}

\newcommand{\ql}{DGQL\xspace}

\usepackage{hyperref}
\hypersetup{
	colorlinks, linkcolor={black},
	citecolor={black}, urlcolor={black},
	pdftitle={MillenniumDB: A Persistent, Open-Source, Graph Database},    
	pdfauthor={IMFD Team}     
}

\newcommand{\objs}{\mathsf{Obj}}
\newcommand{\labels}{\mathcal{L}}
\newcommand{\keys}{\mathcal{K}}
\newcommand{\cprop}{\mathcal{P}}
\newcommand{\values}{\mathcal{V}}
\newcommand{\vars}{\mathsf{Var}}

\def\ojoin{\setbox0=\hbox{$\Join$}\rule[0.1ex]{.25em}{.6pt}\llap{\rule[1ex]{.25em}{.6pt}}}
\def\leftouterjoin{\mathbin{\ojoin\mkern-6.6mu\Join}}
\newcommand{\LOJ}{\mathbin{\leftouterjoin}}

\newcommand{\Dom}[1]{\mathsf{dom}(#1)}
\newcommand{\Var}[1]{\mathsf{var}(#1)}
\newcommand{\inv}[1]{#1^{-}}

\newcommand{\map}{\mu}

\newcommand{\semp}[2]{\llbracket #1\rrbracket_{#2}}

\newcommand{\avar}[1]{\mathsf{AVars}(#1)}
\newcommand{\evar}[1]{\mathsf{EVars}(#1)}
\newcommand{\qvars}[1]{\mathsf{Vars}(#1)}

\newcommand{\lab}{\texttt{lab}}
\newcommand{\prop}{\texttt{prop}}

\lstdefinestyle{ttl}
{	language=SPARQL, 
	basicstyle=\ttfamily\scriptsize, 
	captionpos=b, 
	showstringspaces=false, 
	numbers=none, 
	morekeywords={SIMILARITY, TOP, MINUS, WITHIN, SERVICE, PREFIX, SELF, SELECT, MATCH, WHERE, FILTER}
}

\newcommand{\riveros}[1]{\todo[inline, color=red!30]{\textbf{Riveros}: #1}}

\makeatletter
\newenvironment{customlegend}[1][]{%
    \begingroup
    \pgfplots@init@cleared@structures
    \pgfplotsset{#1}%
}{%
    \pgfplots@createlegend
    \endgroup
}%

\def\addlegendimage{\pgfplots@addlegendimage}
\makeatother

\newcommand{\AAND}{~\operatorname{AND}~}
\newcommand{\OPT}{~\operatorname{OPT}~}

\lstdefinelanguage{dql}{
    keywords = {SELECT,MATCH,WHERE,OPTIONAL,TYPE,AND,OR,ASC,DESC,LIMIT,ORDER,BY}
}

\lstdefinestyle{dqls}{
    inputencoding=utf8,
    basicstyle=\ttfamily\small,
    language=dql,
    basewidth=0.52em, 
    keywordstyle=\textbf,
    upquote=true,
    escapechar=`
}

\newcommand{\dqlt}[1]{\texttt{#1}}
\newcommand{\dqlkw}[1]{\textbf{\dqlt{#1}}}

\lstnewenvironment{dql}[1][]{%
    \noindent\minipage[b]{\linewidth}%
    \centering%
    \tabular{@{}c@{}}%
    \lstset{style=dqls,#1}
}{\endtabular\endminipage\vspace{5pt}}

\usetikzlibrary{arrows,positioning,backgrounds} 
\tikzset{
    ert/.style={ 
     	rt,
     	dashed
     	}, 
    erectw/.style={ 
     	rectw,
     	dashed
     	},
}

\begin{document}

\title[MillenniumDB: A Persistent, Open-Source, Graph Database]{MillenniumDB: A Persistent, Open-Source, Graph Database}

\author[D. Vrgo\v{c}]{Domagoj Vrgo\v{c}}
\affiliation{%
  \institution{PUC Chile \& IMFD}
  \country{Chile}
}
\email{dvrgoc@ing.puc.cl}

\author[C. Rojas]{Carlos Rojas}
\affiliation{%
  \institution{IMFD}
  \country{Chile}
}
\email{cirojas6@uc.cl}

\author[R. Angles]{Renzo Angles}
\affiliation{%
  \institution{University of Talca \& IMFD}
  \country{Chile}
}
\email{rangles@utalca.cl}

\author{Marcelo Arenas}
\affiliation{%
  \institution{PUC Chile \& IMFD}
  \country{Chile}
}
\email{marenas@ing.puc.cl}

\author[D. Arroyuelo]{Diego Arroyuelo}
\affiliation{%
  \institution{UTFSM Chile \& IMFD}
  \country{Chile}
}
\email{darroyue@inf.utfsm.cl}

\author[C. Buil Aranda]{Carlos Buil Aranda}
\affiliation{%
  \institution{UTFSM Chile \& IMFD}
  \country{Chile}
}
\email{cbuil@inf.utfsm.cl}

\author[A. Hogan]{Aidan Hogan}
\affiliation{%
  \institution{DCC, University of Chile \& IMFD}
  \country{Chile}
}
\email{ahogan@dcc.uchile.cl}

\author[G. Navarro]{Gonzalo Navarro}
\affiliation{%
  \institution{DCC, University of Chile \& IMFD}
  \country{Chile}
}
\email{gnavarro@dcc.uchile.cl}

\author[C. Riveros]{Cristian Riveros}
\affiliation{%
  \institution{PUC Chile \& IMFD}
  \country{Chile}
}
\email{cristian.riveros@uc.cl}

\author[J. Romero]{Juan Romero}
\affiliation{%
  \institution{PUC Chile \& IMFD}
  \country{Chile}
}
\email{jaromero6@uc.cl}


\renewcommand{\shortauthors}{Vrgo\v{c} and Rojas, et al.}

\begin{abstract}
In this systems paper, we present MillenniumDB: a novel graph database engine that is modular, persistent, and open source. MillenniumDB is based on a graph data model, which we call domain graphs, that provides a simple abstraction upon which a variety of popular graph models can be supported. The engine itself is founded on a combination of tried and tested techniques from relational data management, state-of-the-art algorithms for worst-case-optimal joins, as well as graph-specific algorithms for evaluating path queries. In this paper, we present the main design principles underlying MillenniumDB, describing the abstract graph model and query semantics supported, the concrete data model and query syntax implemented, as well as the storage, indexing, query planning and query evaluation techniques used. We evaluate MillenniumDB over real-world data and queries from the Wikidata knowledge graph, where we find that it outperforms other popular persistent graph database engines (including both enterprise and open source alternatives) that support similar query features.
\end{abstract}

%
%
\maketitle

\section{Introduction}
\label{sec:intro}
Recent years have seen growing interest in graph databases~\cite{AnglesG08}, wherein nodes represent entities of interest, and edges represent relations between those entities. In comparison with alternative data models, graphs offer a flexible and often more intuitive representation of particular domains~\cite{AnglesABHRV17}. Graphs forgo the need to define a fixed (e.g., relational) schema for the domain upfront, and allow for modeling and querying cyclical relations between entities that are not well-supported in other data models (e.g., tree-based models, such as XML and JSON). Graphs have long been used as an intuitive way to model data in domains such as social networks, transport networks, genealogy, biological networks, etc. Graph databases further enable specific forms of querying, such as path queries that find entities related by arbitrary-length paths in the graph. 
Graph databases have become popular in the context of NoSQL~\cite{Cattell10}, where alternatives to relational databases are sought for specialized scenarios; Linked Data~\cite{HeathB11}, where graph-structured data are published and interlinked on the Web; and more recently Knowledge Graphs~\cite{HoganBC2020}, where diverse data are integrated at large scale under a common graph abstraction.

Alongside this growing interest, recent years have seen a growing number of models, languages, techniques and systems for managing and querying graph databases~\cite{AnglesG08,AnglesABHRV17}. In the context of NoSQL systems, Neo4j~\cite{Webber12}, which uses the query language Cypher~\cite{FrancisGGLLMPRS18}, is a leading graph database system in practice.\footnote{See, e.g., \url{https://db-engines.com/en/ranking/graph+dbms}} Other popular graph database systems include ArangoDB~\cite{ArangoDB}, JanusGraph~\cite{JanusGraph}, OrientDB~\cite{OrientDB}, TigerGraph~\cite{TigerGraph}, etc., which support Gremlin~\cite{Rodriguez15} and other custom graph query languages. We also find graph database systems supporting the RDF data model and SPARQL query language~\cite{AliSYHN21}, including Allegrograph~\cite{AllegroGraph}, Amazon Neptune~\cite{AmazonNeptune}, Blazegraph~\cite{ThompsonPC14}, GraphDB~\cite{BishopKOPTV11}, Jena TDB~\cite{JenaTDB}, Stardog~\cite{Stardog}, Virtuoso~\cite{Erling12}, and (many) more besides~\cite{AliSYHN21}. In summary, there now exist many graph database systems to choose from. 
Within this graph database landscape, however, 
we foresee the need for yet another alternative: an open source graph database that implements state-of-the-art techniques; offers performance, scale and reliability competitive with (or ideally better than) enterprise systems; supports a variety of graph data models and query languages; and is extensible. To the best of our knowledge, no existing open source graph database system combines these features.

For this reason, we have designed and implemented \textit{MillenniumDB}: a persistent graph database system that aims to better align the theory and practice for graph database systems, being based on a solid theoretical foundation and state-of-the-art techniques, while being applicable in practical settings. The overall design and implementation of MillenniumDB is based on the following goals:

\begin{itemize}
\item \textit{Generality}: MillenniumDB aims to support various graph database models and query languages by generalizing them and implementing techniques for the more general setting.
\item \textit{State-of-the-art}: With the goal of reaching state-of-the-art performance, MillenniumDB incorporates and combines techniques from the recent research literature.
\item \textit{Theoretically well-founded}: MillenniumDB provides clear semantics for its underlying interfaces, and prioritizes techniques that provide theoretical guarantees.
\item \textit{Modularity}: In order to enable extensibility, MillenniumDB follows a modular design that decouples a graph database system into components with clearly defined interfaces.
\item \textit{Open source}: The code for MillenniumDB is available under an open source license \cite{millenniumDB}, with the goal that it may be extended and reused by other researchers and practitioners.
\end{itemize}

%

These research goals imply significant engineering and research challenges that form a road-map for the development of MillenniumDB. In this systems paper, we present a first milestone in this development: the first release of MillenniumDB, which establishes the foundations and general architecture of a graph database system that will be extended in the coming years. These foundations include the proposal of a data model -- called \textit{domain graphs} -- that provides a simple abstraction capable of supporting multiple graph models and query languages, a novel query syntax adapted for this model, and a query engine that includes state-of-the-art join and path algorithms. Various benchmarks over the Wikidata knowledge graph~\cite{VrandecicK14} show that this first release of MillenniumDB generally outperforms prominent graph database systems -- namely Blazegraph, Neo4j, Jena and Virtuoso --  in terms of query performance. 

\paragraph{Paper structure} The rest of this paper is structured as follows:
\begin{itemize}
\item In Section~\ref{sec:data}, we discuss graph data models that are popular in current practice, and propose \textit{domain graphs} as an abstraction of these models that we implement in MillenniumDB.
\item In Section~\ref{sec:lang}, we describe the query language of MillenniumDB, and how it takes advantage of \datas.
\item In Section~\ref{sec:architecture}, we explain how MillenniumDB stores data and evaluates queries.
\item In Section~\ref{sec:benchmark}, we provide an experimental evaluation
of the proposed methods on a large body of queries over the Wikidata knowledge graph.

\item In Section~\ref{sec:concl}, we provide some concluding remarks and ideas for future research.
\end{itemize}

\paragraph{Supplementary material} The source code of MillenniumDB is provided in full at \cite{millenniumDB}. Experimental data is given at \cite{benchmark}. We also provide an extended version of the paper at \cite{extended}, with additional details on the query syntax and semantics used in MillenniumDB.

\section{Data model}
\label{sec:data}

In this section, we present the graph data model upon which MillenniumDB is based, and discuss how it generalizes existing graph data models such as RDF and property graphs. We also show its utility in concisely modeling real-world knowledge graphs that contain higher-arity relations, such as Wikidata \cite{VrandecicK14}.

\subsection{Domain Graphs}
\label{sec-proposal}

The structure of knowledge graphs is captured in MillenniumDB via \textit{\datas}, which follow the natural idea of assigning ids to directed labeled edges in capture higher-arity relations within graphs~\cite{HernandezHK15,IlievskiGCDYRLL20,LassilaSBBBKKLST}. Formally, assume a universe $\objs$ of objects (ids, strings, numbers, IRIs, etc.). We define \datas as follows:

\begin{definition}
A \emph{domain graph} $G = (O,\gamma)$ consists of a finite set of objects $O\subseteq \objs$ and a partial mapping $\gamma : O \rightarrow O \times O \times O$. 
\end{definition}

Intuitively, $O$ is the set of objects that appear in our graph database, and $\gamma$ models edges between objects. If $\gamma(e)=(n_1,t,n_2)$, this states that the edge $(n_1,t,n_2)$ has id $e$, type $t$, and links the source node $n_1$ to the target node $n_2$.\footnote{Herein, we say ``\textit{edge type}'' rather than ``\textit{edge label}'' to highlight that the type forms part of the edge, rather than being an annotation on the edge, as in property graphs.} We can analogously define our model as a relation:
\[\textsc{DomainGraph}(\textsf{source},\textsf{type},\textsf{target},\underline{\textsf{eid}}) \]
where \underline{\textsf{eid}} (edge id) is a primary key of the relation. 

The \data model of MillenniumDB already subsumes the RDF graph model~\cite{CyganiakWL14}. Recall that an RDF graph is a set of triples of the form $(a,b,c)$. An RDF graph can be visualized as a \textit{directed labeled graph}~\cite{AnglesABHRV17,MendelzonW89}, which is a set of edges of the form \gedge[arrin][0.6cm]{$a$}{$b$}{$c$}, where $a$ is the source node (aka \textit{subject}), $b$ the edge type (aka \textit{predicate}), and $c$ the target node (aka \textit{object}). Alternatively, an RDF graph can be seen as a relation $\textsc{RDFGraph}(\textsf{source},\textsf{type},\textsf{target})$. To show how RDF is modeled in \datas, consider the following edge, claiming that Michelle Bachelet was the president of Chile.

\medskip
\begin{center}
\gedge[arrin][2.5cm]{Michelle Bachelet}{position held}{President of Chile}
\end{center}
\medskip

\noindent
We can encode this triple in a \data by storing the tuple \textsf{(Michelle Bachelet, position held, President of Chile, e)} in the \textsf{DomainGraph} relation, where \textsf{e} denotes a unique (potentially auto-generated) edge id, or equivalently stating that:
$$\gamma(\textsf{e}) = \textsf{(Michelle Bachelet, position held, President of Chile)}.$$

\noindent
The id of the edge itself is not accessible in the RDF data model. However, the edge id can be useful when modeling RDF* graphs or named graphs based on RDF, as we will discuss in Section~\ref{ssec:whydg}.



\begin{figure}
\begin{tikzpicture}
\node[nrect] (n1) {
\alt{
\uri{gender} & = \uri{"female"}\\[-0.7em]
\uri{children} & = \uri{"3"}\\[-0.7em]
\uri{first name} & = \uri{"Michelle"}\\[-0.7em]
\uri{last name} & = \uri{"Bachelet"}\\[-0.2em] 
}};

\node[rt] (ln1) at (n1.north) {$n_1$ $:$ \uri{human}};

\node[rect, right=2.4cm of n1] (n2) {
\alt{ 
\uri{gender} & = \uri{"male"}\\[-0.7em]
\uri{children} & = \uri{"2"}\\[-0.7em]
\uri{first name} & = \uri{"Alberto"}\\[-0.7em]
\uri{last name} & = \uri{"Bachelet"}\\[-0.7em]
\uri{death} & = \uri{"12 March 1974"}\\[-0.2em]
}};

\node[rt] (ln2) at (n2.north) {$n_2$ $:$ \uri{human}};

\draw[arrout,bend left=7,transform canvas={yshift=0.25cm}] (n1) to 
node[lab] (le1) {$e_1$ $:$ \uri{father}} 
(n2);

\draw[arrout,bend left=7,transform canvas={yshift=0.27cm}] (n2) to 
node[lab] (le2) {$e_2$ $:$ \uri{child}}
(n1);

\node[erect,anchor=south] (e2) at (n2.south-|le2.mid) {
\alt{
\uri{order} & = \uri{"2"}\\[-0.2em]
}
};
\end{tikzpicture}
\caption{A property graph with two nodes and two edges. We use the order property on edge $e_2$ to indicate that Michelle Bachelet is the second child of Alberto Bachelet}\label{fig:pgmb}
\end{figure}
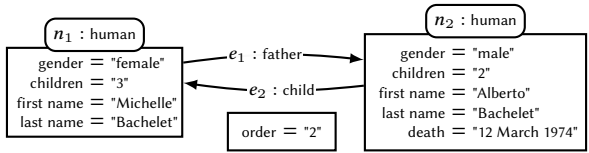

Property graphs~\cite{AnglesABHRV17} further allow nodes and edges to be annotated with labels and property--value pairs, as shown in Figure~\ref{fig:pgmb}. Domain graphs can directly capture property graphs, where, for example, the property--value pair $(\textsf{gender},\textsf{"female"})$ on node $n_1$ can be represented by an edge $\gamma(e_3) = (n_1,\textsf{gender},\textsf{female})$, the property--value pair $(\textsf{order},\textsf{"2"})$ on an edge $e_2$ becomes $\gamma(e_4) = (e_2,\textsf{order},\textsf{2})$, the label on $n_1$ becomes $\gamma(e_5) = (n_1,\textsf{label},\textsf{human})$, the label on $e_1$ becomes the type of the edge $\gamma(e_1) = (n_1,\textsf{father},n_2)$, etc.\footnote{To represent edges in property graphs that permit multiple labels, multiple edges with different types can be added (or the labels can be added on the edge ids).} However, given a legacy property graph, there are some potential ``incompatibilities'' with the resulting domain graph; for example, strings like \textsf{"male"}, labels like \textsf{human}, etc., now become nodes in the graph, generating new paths through them that may affect query results.

For stricter backwards compatibility with legacy property graphs (where desired), MillenniumDB implements a simple extension of the \data model, called \textit{property \datas}, which allows for \textit{external annotation}, i.e., adding labels and property--value pairs to nodes and edges without creating new nodes and edges. Formally, if $\labels$ is a set of labels, $\cprop$ a set of properties, and  $\values$ a set of values, we define a property \data as follows:

\begin{definition}
 A \emph{property \data} is defined as a tuple $G=(O,\gamma,\lab,\prop)$, where:
\begin{itemize}
    \item $(O,\gamma)$ is a \data;
    \item $\lab : O \rightarrow 2^\labels$ is a function
    assigning a finite set of labels to an object; and
    \item $\prop : O \times \cprop \rightarrow \values$ is a partial function assigning a value to a certain property of an object. 
\end{itemize}
Moreover, we assume that for each object $o \in O$, there exists a finite number of properties $p \in \cprop$ such that $\prop(o,p)$ is defined.
\end{definition}






In summary, \datas capture the graph structure of our model, while property \datas additionally permit annotating that graph structure by adding labels and property--value pairs to objects, as ordinary property graphs do; such annotations are then considered external from the graph structure. The property graph in Figure~\ref{fig:pgmb}
can be represented with the following property \data $G=(O,\gamma,\lab,\prop)$, where the graph structure is as follows:
{\small
\begin{align*}
O & = \{n_1,n_2,e_1,e_2\}, \\
\gamma(e_1) & = (n_1,\textsf{father},n_2),\\
\gamma(e_2) & = (n_2,\textsf{child},n_1),
\end{align*}}
and the annotations of the graph structure are as follows:
\vspace{-2mm}
\begin{center}
{\small
\begin{tabular}{ll}
$\textsf{lab}(n_1)= \textsf{human}$ &
$\textsf{prop}(n_1,\textsf{last name}) = \textsf{"Bachelet"}$\\
$\textsf{lab}(n_2) = \textsf{human}$ &
$\textsf{prop}(n_2,\textsf{gender}) = \textsf{"male"}$\\
$\textsf{prop}(e_2,\textsf{order}) = \textsf{"2"}$ &
$\textsf{prop}(n_2,\textsf{children}) = \textsf{"2"}$\\
$\textsf{prop}(n_1,\textsf{gender}) = \textsf{"female"}$ &
$\textsf{prop}(n_2,\textsf{first name}) = \textsf{"Alberto"}$\\
$\textsf{prop}(n_1,\textsf{children}) = \textsf{"3"}$ &
$\textsf{prop}(n_2,\textsf{last name}) = \textsf{"Bachelet"}$\\
$\textsf{prop}(n_1,\textsf{first name}) = \textsf{"Michelle"}$ &
$\textsf{prop}(n_2,\textsf{death}) = \textsf{"12 March 1974"}$
\end{tabular}
}
\end{center}
%
\medskip

\noindent The relational representation of property \data then adds two new relations alongside \textsc{DomainGraph}:
\begin{center}
$\textsc{Labels}(\textsf{object},\textsf{label})$,\\
$\textsc{Properties}(\underline{\textsf{object},\textsf{property}},\textsf{value}),$
\end{center}
where \underline{\textsf{object},\,\textsf{property}} is a primary key of the second relation, with the first relation allowing multiple labels per object.

\subsection{Why \datas?}\label{ssec:whydg}

Why did we choose \datas as the model of MillenniumDB? As discussed in the previous section, it can be used to model both directed labeled graphs (like RDF) as well as property graphs. It also has a natural relational expression, which facilitates its implementation in a query engine. But it is also heavily inspired by the needs of real-world knowledge graphs like Wikidata~\cite{VrandecicK14}.

Consider the two Wikidata statements shown in Figure~\ref{fig:mb}. Both statements claim that Michelle Bachelet was a president of Chile, and both are associated with nested \textit{qualifiers} that provide additional information: in this case a start date, an end date, who replaced her, and whom she was replaced by. There are two statements, indicating two distinct periods when she held the position. Also the ids for objects (for example, \textsf{Q320} and \textsf{P39}) are shown; any positional element can have an id and be viewed as a node in the knowledge graph. 

\newcommand{\wid}[1]{{\color{gray}~[#1]}} 

\begin{figure}[t]
\centering
{\sf
\begin{tabular}{l}
\large\textbf{Michelle Bachelet\wid{Q320}}\\[1ex]
\large\qquad position held\wid{P39} \quad President of Chile\wid{Q466956}\\
\qquad \qquad 
{\begin{tabular}{l@{\qquad~~}l}
start date \wid{P580} & 2014-03-11\\
end date \wid{P582} 	& 2018-03-11\\
replaces\wid{P155}	& Sebastián Piñera\wid{Q306}\\
replaced by\wid{P156} & Sebastián Piñera\wid{Q306}\\
\end{tabular}}\\
\\[-1ex]
\large\qquad position held\wid{P39} \quad President of Chile\wid{Q466956}\\
\qquad \qquad {\begin{tabular}{l@{\qquad~~}l}
start date \wid{P580} & 2006-03-11\\
end date \wid{P582} & 2010-03-11\\
replaces\wid{P155} & Ricardo Lagos\wid{Q331}\\
replaced by\wid{P156} & Sebastián Piñera\wid{Q306}\\
\end{tabular}}
\end{tabular}
}
\caption{Wikidata statement group for Michelle Bachelet \label{fig:mb}}
\end{figure}

\begin{figure}[t]
\setlength{\vgap}{0.45cm}
\setlength{\hgap}{2.5cm}
\centering
\begin{tikzpicture}
\node[iri] (e1) {$e_1$};

\node[iri,right=\hgap of e1] (pe1) {position held}
  edge[arrin] node[lab] {type} (e1);
  
\node[iri,above=\vgap of pe1] (se1) {Michelle Bachelet}
  edge[arrin] node[lab] {source} (e1);
  
\node[iri,below=\vgap of pe1] (oe1) {President of Chile}
  edge[arrin] node[lab] {target} (e1);
 
\node[iri,above=\vgap of se1] (sp) {Sebastian Piñera}
  edge[arrin] node[lab] {replaced by} (e1)
  edge[arrin,bend right=19] node[lab,pos=0.45] {replaces} (e1);
  
\node[iri,right=\hgap of pe1] (e2) {$e_2$}
  edge[arrout] node[lab] {type} (pe1)
  edge[arrout] node[lab] {source} (se1)
  edge[arrout] node[lab] {target} (oe1)
  edge[arrout] node[lab] {replaced by} (sp);
  
\node[iri,below=2.2\vgap of e1] (sd1) {2014-03-11}
  edge[arrin] node[lab] {start date} (e1);
\node[iri,above=2.2\vgap of e1] (ed1) {2018-03-11}
  edge[arrin] node[lab] {end date} (e1);

\node[iri,below=2.2\vgap of e2] (sd2) {2006-03-11}
  edge[arrin] node[lab] {start date} (e2);
\node[iri,above=2.2\vgap of e2] (ed2) {2010-03-11}
  edge[arrin] node[lab] {end date} (e2);
  
\node[iri,below=\vgap of oe1] (rl) {Ricardo Lagos}
  edge[arrin,bend right=17] node[lab] {replaces} (e2);
\end{tikzpicture}
\caption{Directed labeled graph reifying the statements of Figure~\ref{fig:mb} \label{fig:delg}}
\end{figure}
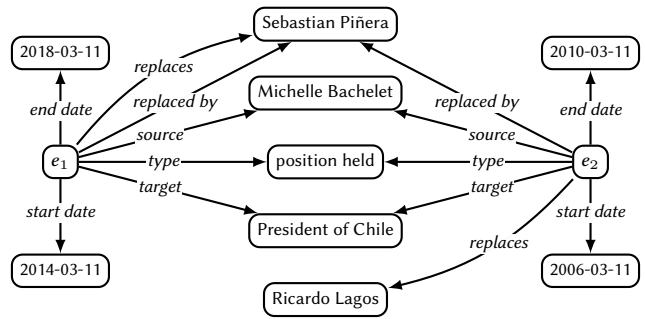

Representing statements like this in a directed labeled graph requires some form of \textit{reification} to decompose $n$-ary relations into binary relations~\cite{HernandezHK15}. For example, Figure~\ref{fig:delg} shows a graph where $e_1$ and $e_2$ are nodes representing $n$-ary relationships. The reification is given by the use of the edges typed as \textsf{source}, \textsf{type} and \textsf{target}. As before, we use human-readable nodes and labels, where in practice, a node \gnode{Sebastián Piñera} will rather be given as the identifier \gnode{Q306}, and an edge type \textsf{replaces} will rather be given as \textsf{P155}.


A number of graph models have been proposed to capture higher-arity relations more concisely, including property graphs~\cite{FrancisGGLLMPRS18} and RDF*~\cite{Hartig17}. However, both have limitations that render them incapable of modeling the statements shown in Figure~\ref{fig:mb} without resorting to reification~\cite{HoganRRS19}. On the one hand, property graphs allow labels and property--value pairs to be associated with both nodes and edges. For example, the statements of Figure~\ref{fig:mb} can be represented as the property graph:

\begin{center}
\vspace{2pt}
\setlength{\vgap}{1.5cm}
\setlength{\hgap}{1.5cm}
\begin{tikzpicture}
  \node[nrect] (n1) {
   \alt{
       \uri{name} & =\,\uri{"Michelle Bachelet"} } };
  \node[rt] (ln1) at (n1.north) {$n_1$ : \uri{human}};
  
  \node[nrect,right=\hgap of n1] (n2) {
   \alt{
      \uri{name} & =\,\uri{"President of Chile"} } };
  \node[rt] (ln2) at (n2.north) {$n_2$ : \uri{public office}};

  \draw[arrout,pos=0.5,bend left=25] (ln1) to node[rte,yshift=-0.1cm] (le1)
   {$e_1$ $:$ position held}
  (ln2);
  \node[erect,anchor=south,yshift=-0.5ex] at (le1.north) (e1) {
     \alt{ \uri{start date} & =\,\uri{"11 March 2014"}\\[-0.7em]
           \uri{end date} & =\,\uri{"11 March 2018"}\\[-0.7em]
           \uri{replaces} & =\,\uri{"Sebastián Piñera"}\\[-0.7em]
           \uri{replaced by} & =\,\uri{"Sebastián Piñera"} } };
  
  \draw[arrout,pos=0.5,bend right=25] (n1) to node[rte,yshift=-0.1cm] (le2) {$e_2$ $:$ position held} (n2);
  \node[erect,anchor=north,yshift=0.5ex] (e2) at (le2.south) {
     \alt{ \uri{start date} & =\,\uri{"11 March 2006"}\\[-0.7em]
           \uri{end date} & =\,\uri{"11 March 2010"}\\[-0.7em]
           \uri{replaces} & =\,\uri{"Ricardo Lagos"}\\[-0.7em]
           \uri{replaced by} & =\,\uri{"Sebastián Piñera"}\\ } };
\end{tikzpicture}
\vspace{2pt}
\end{center}

Though more concise than reification, labels, properties and values are considered to be simple strings, which are disjoint with nodes; for example, \textsf{"Ricardo Lagos"} is neither a node nor a pointer to a node, but a string, which would complicate, for example, querying for the parties of presidents that Michelle Bachelet replaced. 

%
%
%

On the other hand, RDF* allows an edge to be a node. For example, the first statement of Figure~\ref{fig:mb} can be represented as follows in RDF*:

\begin{center}
\setlength{\vgap}{0.8cm}
\setlength{\hgap}{1.9cm}
\vspace{2pt}
\begin{tikzpicture}
\node[iri,anchor=center,minimum width=6.3cm,minimum height=0.7cm,enode] (e1) {};
\node[iri,right=1ex of e1.west,anchor=west] (sp1) {Michelle Bachelet};
\node[iri,right=\hgap of sp1] (ch1) {President of Chile}
  edge[arrin] node[lab] {position held} (sp1);
  
\node[iri,below=\vgap of e1] (mb) {Sebastián Piñera}
  edge[arrin,bend right=40] node[lab,xshift=1ex] {replaces} (e1)
  edge[arrin,bend left=40] node[lab,xshift=-1ex] {replaced by} (e1);

\node[iri,left=\hgap of mb] (ttt) {2014-03-11}
  edge[arrin] node[lab] {start date} (e1);

\node[iri,right=\hgap of mb] (ttf) {2018-03-11}
  edge[arrin] node[lab] {end date} (e1);
\end{tikzpicture}
\vspace{2pt}
\end{center}
However,  we can only represent one of the statements (without reification), as we can only have one distinct node per edge; if we add the qualifiers for both statements, then we would not know which start date pairs with which end date, for example.

%
%
%

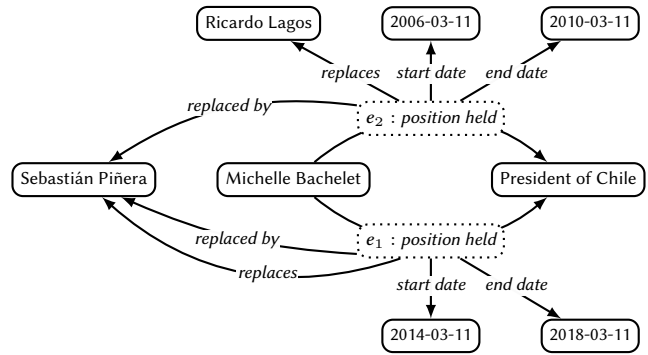
\begin{figure}[t]
\setlength{\vgap}{0.8cm}
\setlength{\hgap}{1.6cm}
\centering
\begin{tikzpicture}
\node[iri,anchor=west,yshift=1ex] (sp1) {Michelle Bachelet};
\node[iri,right=\hgap of sp1] (ch1) {President of Chile}
  edge[arrin,bend left=40] node[enode] (e1) {$e_1:$ position held} (sp1)
  edge[arrin,bend right=40] node[enode] (e2) {$e_2:$  position held} (sp1);

\node[iri,below=\vgap of e1] (ttt) {2014-03-11}
  edge[arrin] node[lab] {start date} (e1);
  
\node[iri,left=0.5\hgap of sp1] (mb) {Sebastián Piñera}
  edge[arrin,bend right=30] node[lab,xshift=1ex] {replaces} (e1.210)
  edge[arrin,bend right=10] node[lab,xshift=-1ex] {replaced by} (e1.190)
  edge[arrin,bend left=20] node[lab,xshift=-1ex] {replaced by} (e2.170);

\node[iri,right=0.5\hgap of ttt] (ttf) {2018-03-11}
  edge[arrin] node[lab,xshift=1ex] {end date} (e1);

\node[iri,above=\vgap of e2] (ttt2) {2006-03-11}
  edge[arrin] node[lab] {start date} (e2);
  
\node[iri,left=0.5\hgap of ttt2] (rl) {Ricardo Lagos}
  edge[arrin] node[lab] {replaces} (e2);

\node[iri,right=0.5\hgap of ttt2] (ttf2) {2010-03-11}
  edge[arrin] node[lab,xshift=1ex] {end date} (e2);


\end{tikzpicture}
\caption{Domain graph for Figure~\ref{fig:mb} \label{fig:dg}}
\end{figure}

The \data model allows us to capture higher-arity relations more directly. In Figure \ref{fig:dg} we have one possible representation of the statements from Figure \ref{fig:mb}. We only show edge ids as needed (all edges have ids). We do not use the ``property part" of our data model for external annotation, considering that the elements of Wikidata statements shown can form nodes in the graph itself.


Domain graphs are inspired by \textit{named graphs} in RDF/SPARQL (where named graphs have one triple/edge each). Both domain graphs and named graphs can be represented as quads. However, the edge ids of domain graphs identify each quad, which, as we will discuss in Section~\ref{sec:architecture}, simplifies indexing. Named graphs were proposed to represent multiple RDF graphs for publishing and querying. SPARQL thus uses syntax -- such as \textsf{FROM}, \textsf{FROM NAMED}, \textsf{GRAPH} -- that is unintuitive when querying singleton named graphs representing higher-arity relations. Moreover, SPARQL does not support querying paths that span named graphs; in order to support path queries over singleton named graphs, all edges would need to be duplicated (virtually or physically) into a single graph~\cite{HernandezHK15}. Named graphs (with multiple edges) could be supported in domain graphs using a reserved term \textsf{graph}, and edges of the form $\gamma(e_3) = (e_1,\textsf{graph},g_1)$, $\gamma(e_4) =(e_2,\textsf{graph},g_1)$; optionally, \textit{named domain graphs} could be considered in the future to support multiple domain graphs with quins.

The idea of assigning ids to edges/triples for similar purposes as described here is a natural one, and not new to this work. \citet{HernandezHK15} explored using singleton named graphs in order to represent Wikidata qualifiers, placing one triple in each named graph, such that the name acts as an id for the triple. In parallel with our work, recently a data model analogous to domain graphs has been independently proposed for use in Amazon Neptune, which the authors call 1G~\cite{LassilaSBBBKKLST}. Their proposal does not discuss a formal definition for the model, nor a query language, storage and indexing, implementation, etc., but the reasoning and justification that they put forward for the model is similar to ours. To the best of our knowledge, our work is the first to describe a query language, storage and indexing schemes, query planner -- and ultimately a fully-fledged graph database engine -- built specifically for this model. Furthermore, with property domain graphs, we support annotation external to the graph, which we believe to be a useful extension that enables better compatibility with property graphs.

Table~\ref{tab:gmodel} summarizes the features that are directly supported by the respective graph models themselves without requiring \textit{reserved terms}, which would include, for example, \textsf{source}, \textsf{label} and \textsf{target} in Figure~\ref{fig:delg} (all features except \textit{External annotation} can be supported in all models with reserved vocabulary). Reserved terms can add indirection to modeling (e.g., reification~\cite{HernandezHK15}), and can clutter the data, necessitating more tuples or higher-arity tuples to store, leading to more joins and/or index permutations. The features are then defined as follows, considering directed (labeled) edges:

\begin{itemize}
\item \textit{Edge type/label}: assign a type or label to an edge.
\item \textit{Node label}: assign labels to nodes.
\item \textit{Edge annotation}: assign property--value pairs to an edge.
\item \textit{Node annotation}: assign property--value pairs to a node.
\item \textit{External annotation}: nodes/edges can be annotated without adding new nodes or edges.
\item \textit{Edge as node}: an edge can be referenced as a node (this allows edges to be connected to nodes of the graph).
\item \textit{Edge as nodes}: a single unique edge can be referenced as multiple nodes.
\item \textit{Nested edge nodes}: an edge involving an edge node can itself be referenced as a node, and so on, recursively.
\item \textit{Graph as node}: a graph can be referenced as a node.
\end{itemize}

Some \nmark's in Table~\ref{tab:gmodel} are more benign than others; for example, \textit{Node label} requires a reserved term (e.g., \textsf{rdf:type}), but no extra tuples; on the other hand, \textit{Edge as node} requires reification, using at least one extra tuple, and at least one reserved term.

Wikidata requires \textit{Edge as nodes} for cases as shown in Figure~\ref{fig:mb} (since values such as \textsf{Ricardo Lagos} are themselves nodes). Only named graphs, domain graphs and property domain graphs can model such examples without reserved terms. Comparing named graphs and domain graphs, the latter sacrifices the ``\textit{Graph as node}'' feature without reserved vocabulary to reduce indexing permutations (discussed in Section~\ref{sec:architecture}). Such a feature is not needed by Wikidata. Property domain graphs further support external annotation, and thus better compatibility with legacy property graphs.


\begin{table}
\caption{The features supported by graph models without reserved terms (NG = Named Graphs, PG = Property Graphs, DG = Domain Graphs, PDG = Property Domain Graphs) \label{tab:gmodel}}
\begin{tabular}{ccccccc}
\toprule
 & RDF & RDF* & NG & PG & DG & PDG \\
\toprule
\textit{Edge type/label} & \ymark & \ymark & \ymark & \ymark & \ymark & \ymark \\
\textit{Node label} & \nmark & \nmark & \nmark & \ymark & \nmark & \ymark \\
\textit{Edge annotation} & \nmark & \ymark & \ymark & \ymark & \ymark & \ymark \\
\textit{Node annotation} & \ymark & \ymark & \ymark & \ymark & \ymark & \ymark \\
\textit{External annotation} & \nmark & \nmark & \nmark & \ymark & \nmark & \ymark \\
\textit{Edge as node} & \nmark & \ymark & \ymark & \nmark & \ymark & \ymark \\
\textit{Edge as nodes} & \nmark & \nmark & \ymark & \nmark & \ymark & \ymark \\
\textit{Nested edge nodes} & \nmark & \ymark & \ymark & \nmark & \ymark & \ymark \\
\textit{Graph as node} & \nmark & \nmark & \ymark & \nmark & \nmark & \nmark \\
\bottomrule
\end{tabular}
\end{table}

\section{Query language}
\label{sec:lang}

We now discuss the query language that MillenniumDB currently supports. We first introduce the common primitives supported by graph query languages, and then discuss the syntax and semantics of a concrete query language currently supported by our system.

\subsection{Features of Graph Query Languages}

A variety of query languages for graphs have been proposed down through the years~\cite{AnglesG08}, including AQL~\cite{ArangoDB} (used by ArangoDB), Cypher~\cite{FrancisGGLLMPRS18} (used by Neo4j~\cite{Webber12}), Gremlin~\cite{Rodriguez15} (used by Amazon Neptune~\cite{AmazonNeptune}, JanusGraph~\cite{JanusGraph}, Neo4j~\cite{Webber12}, OrientDB~\cite{OrientDB}), G-Core~\cite{AnglesABBFGLPPS18}, GSQL~\cite{TigerGraph} (used by TigerGraph~\cite{TigerGraph}) SPARQL~\cite{HarrisS13} (used by various systems~\cite{AliSYHN21}, including Allegrograph~\cite{AllegroGraph}, Amazon Neptune~\cite{AmazonNeptune}, Blazegraph~\cite{ThompsonPC14}, GraphDB~\cite{BishopKOPTV11}, Jena TDB~\cite{JenaTDB}, Stardog~\cite{Stardog} and Virtuoso~\cite{Erling12}), SQL-Match~\cite{OrientDB} (used by OrientDB~\cite{OrientDB}).

While the syntax of these query languages vary widely, they share a large common core in terms of querying primitives. Specifically, all such query languages support different types of \textit{graph patterns} that transform a graph into a relation (aka.\ a table) of results~\cite{AnglesABHRV17}. The simplest form of graph pattern is a \textit{basic graph pattern}, which in its most general form, is a graph pattern structured like the data, but that also allows variables to replace constants. Next we have \textit{navigational graph patterns}, which allow for specifying path expressions that can match arbitrary-length paths in the graphs. We also have \textit{relational graph patterns} which, noting that a graph pattern converts a graph into a relation, allows the use of relational operators to transform and/or combine the results of one or more graph patterns into a single relation. Finally, query languages may support other features, such as \textit{solution modifiers} that order or limit results, convert results back into a graph, etc.

\subsection{Domain Graph Queries}

Per our goal of supporting multiple graph models, MillenniumDB aims to support a number of graph query languages, such as those that we previously discussed. However, no existing query language would take full advantage of the property domain graph model defined in the previous section. We have thus implemented a base query language, called \ql, which closely resembles Cypher~\cite{FrancisGGLLMPRS18}, but is adapted for the property domain graph model, and adds features of other query languages, such as SPARQL, that are commonly used for querying knowledge graphs like Wikidata~\cite{BonifatiMT20}. Herein we provide a guided tour of the syntax of \ql.

A \ql query takes the following high-level form:

\begin{dql}
SELECT Variables
MATCH Pattern
WHERE Filters
\end{dql}

\noindent
When evaluated over a property \data, such a query will return a multiset of mappings binding \dqlt{Variables} to database objects (or values) that satisfy the \dqlt{Pattern} specified in the \dqlkw{MATCH} clause and the \dqlt{Filters} specified in the \dqlkw{WHERE} clause.


\paragraph*{Querying objects.} The most basic query will return all the objects (or more precisely, their ids) in our property \data. In MillenniumDB we can achieve this via the following query:

\begin{dql}
SELECT ?x
MATCH (?x)
\end{dql}

\noindent
Of course, one usually wants to select objects with a certain label, or a certain value in a specific property. For instance, if we want to select all people in the property (domain) graph from Figure~\ref{fig:pgmb} with two children, we could do this as follows:

\begin{dql}
SELECT ?x, ?x.gender
MATCH (?x :human { children : "2" })
\end{dql}

\noindent This would return the ids of nodes with label \dqlt{human} and value \dqlt{"2"} for the property \dqlt{children}, along with their associated value for the property \dqlt{gender}, i.e., a single result $\{ \{ \dqlt{?x} \mapsto n_2, \dqlt{?x.gender} \mapsto \dqlt{"male"} \} \}$. 
If $n_2$ did not have a gender, we would still want to return the node that makes the match, but signal that the gender is not specified. To do so, we would return  $\{ \{ \dqlt{?x} \mapsto n_2, \dqlt{?x.gender} \mapsto \dqlt{null} \} \}$.


If we wish to specify a range, we can rather use the \dqlkw{WHERE} clause:

\begin{dql}
SELECT ?x, ?x.gender
MATCH (?x :human)
WHERE ?x.children >= "2"
\end{dql}

\noindent which returns two solutions:  $\{ \{ \dqlt{?x} \mapsto n_2, \dqlt{?x.gender} \mapsto \dqlt{"male"} \},$ $\{ \dqlt{?x} \mapsto n_1, \dqlt{?x.gender} \mapsto \dqlt{"female"} \} \}$. If we replaced \dqlt{>=} with \dqlt{==}, the results would be the same as for the previous query.


%
%
%
%
%

\paragraph*{Querying edges.} In order to query over edges, we can write the following query, which returns $\gamma$, i.e., the relation $\textsc{DomainGraph}$:

\begin{dql}
SELECT *
MATCH (?x)-[?e TYPE(?t)]->(?y)
\end{dql} 

\noindent
The \dqlkw{SELECT} \dqlt{*} operation projects all variables specified in the \dqlkw{MATCH} pattern, while the construct \texttt{(?x)-[?e TYPE(?t)]->(?y)} specifies that we want to connect the object in \texttt{?x} with an object in \texttt{?y}, via an edge with type \texttt{?t} and id \texttt{?e}. This is akin to a query \textsc{DomainGraph}(\texttt{?x,?t,?y,?e}) over the domain graph relation. 

We can also restrict which edges are matched. The following query in \ql will return ids for edges of type child, where both nodes have the same last name, and the child is not the oldest; it also returns the order of the child:

\begin{dql}
SELECT ?e, ?e.order
MATCH (?x)-[?e child]->(?y)
WHERE (?x.lastname == ?y.lastname) AND (?e.order > "1")
\end{dql}

\noindent
This returns $\{ \{ \dqlt{?e} \mapsto e_2, \dqlt{?e.order} \mapsto \dqlt{2} \} \}$ when evaluated over the graph in Figure~\ref{fig:pgmb}. If we need to query for the second oldest (an equality condition), we could use the syntax \dqlt{(?x)-[?e child \{order : "2"\}]->(?y)} on the edge (or replace \dqlt{(?e.order > "1")} with \dqlt{(?e.order == "1")}). As shown here, the \dqlkw{WHERE} clause may use Boolean combinations. Notice that edge types are written ``as is'', which is in contrast to object labels that are prefixed with the \texttt{:} symbol.


\paragraph*{Querying known objects.} Knowing the id of an object, we might want to query more about it. 
%
%
%
For instance, in the domain graph of Figure~\ref{fig:dg} representing statements from Wikidata, we may query for the positions held by Michelle Bachelet as follows:

\begin{dql}
SELECT ?x
MATCH (Michelle Bachelet)-[position held]->(?x)
\end{dql}

\noindent This would return duplicate results $\{ \{ \dqlt{?x} \mapsto \dqlt{President of Chile} \},$  $\{ \dqlt{?x} \mapsto \dqlt{President of Chile} \} \}$ due to the two edges $e_1$ and $e_2$. Recall that for clarity we use human-readable ids, where for Wikidata, the node id \dqlt{Michelle Bachelet} may rather be given as \dqlt{Q320}, and \dqlt{position held} by \dqlt{P39}.

\paragraph*{Path queries.} A key feature of graph databases is their ability to explore paths of arbitrary length. Like SPARQL~\cite{HarrisS13} (but not Cypher\footnote{Of these 2RPQ features, Cypher only supports Kleene star. However, 2RPQs are widely used for querying Wikidata~\cite{BonifatiMT20}, and thus we have prioritized adding this feature. We do not support SPARQL's negated property sets, which are used infrequently~\cite{BonifatiMT20}.}), \ql supports two-way regular path queries (2RPQs), which specify regular expressions over edge types, including concatenation (\dqlt{/}), disjunction (\dqlt{|}), inverses (\verb|^|), optional (\dqlt{?}), Kleene star (\dqlt{*}) and Kleene plus (\dqlt{+}). For instance, if we need to find all descendants of people named Alberto, we could use the regular expression \texttt{child+} in the following way:

\begin{dql}
SELECT ?y
MATCH (?x { first name : "Alberto" })=[child+]=>(?y)
\end{dql}


\noindent This returns $\{ \{ \dqlt{?y} \mapsto n_1  \}\}$ over the graph in Figure~\ref{fig:pgmb}, and would include children of $n_1$, and their children, recursively, if present. We use \dqlt{=[]=>} (rather than \dqlt{-[]->}) to signal a path query in \ql. If we want to capture more results, we could replace \dqlt{[child+]} with an extended expression like \verb#[(child|^father|^mother)+]# which also traverses backwards along edges of type \dqlt{father} and \dqlt{mother}, and may thus return more results. We can also concatenate paths with $\dqlt{/}$, where the expression \verb#[(father|mother)*/sister]# could find sisters, aunts, grand-aunts, and so forth.


Like Cypher (but not SPARQL), \ql can return a single shortest path witnessing the query result via the following construct:

\begin{dql}
SELECT ?y, ?p
MATCH (?x { first name : Alberto })=[?p child+]=>(?y)
\end{dql}

\noindent
The variable \texttt{?p} stores a shortest path connecting each \texttt{Alberto} with each descendant \texttt{?y}. No manipulation of path variables, apart from outputting the result, is currently supported in MillenniumDB, but a full path algebra will be supported in future versions.

\paragraph*{Basic and navigational graph patterns.} Basic graph patterns~\cite{AnglesABHRV17} lie at the core of many graph query languages, including \ql. These are graphs following the same structure as the data model, but where variables are allowed in any position. They can also be seen as expressing natural (multi)joins over sets of atomic edge patterns. In \ql, they are given in the \dqlkw{MATCH} clause. As a way to illustrate this, the following query finds pairs of nodes who share the same father:

\begin{dql}
SELECT ?x, ?y
MATCH (?x)-[father]->(?z),
      (?y)-[father]->(?z)
\end{dql}

\noindent
We evaluate basic graph patterns under \textit{homomorphism-based semantics}~\cite{AnglesABHRV17}, which allows multiple variables in a single result to map to the same element of the data. If we evaluate this query over Figure~\ref{fig:pgmb}, we would thus get one solution: $\{ \{ \dqlt{?x} \mapsto n_1, \dqlt{?y} \mapsto n_1,  \} \}$.

If we further allow path queries within basic graph patterns, we arrive at \textit{navigational graph patterns}~\cite{AnglesABHRV17}. Suppose we need to find, for example, pairs of nodes that share a common ancestor along the paternal line and held the same position; we could write this as:

\begin{dql}
SELECT ?x, ?y
MATCH (?x)=[father+]=>(?z),
      (?y)=[father+]=>(?z),
      (?x)-[position held]->(?w)
      (?y)-[position held]->(?w)
WHERE ?x != ?y
\end{dql}

\noindent This time we filter results that map \dqlt{?x} and \dqlt{?y} to the same node.


\paragraph*{Taking advantage of the \data model.} The \ql query language allows us take full advantage of
domain graphs, by allowing joins between edges, types, etc. To illustrate this, consider the following query, evaluated over the domain graph of Figure~\ref{fig:dg}:

\begin{dql}
SELECT ?x, ?d
MATCH (Michelle Bachelet)-[?e position held]->
                             (President of Chile),
      (?e)-[replaces]->(?x),
      (?e)-[start date]->(?d)
\end{dql}

\noindent The variable \dqlt{?e} invokes a join between an edge and a node, returning two solutions: $\{ \{ \dqlt{?x} \mapsto \dqlt{Ricardo Lagos}, \dqlt{?d} \mapsto \dqlt{2006-03-11} \},$ $\{ \dqlt{?x} \mapsto \dqlt{Sebastián Piñera} , \dqlt{?d} \mapsto \dqlt{2014-03-11} \} \}$. Hence, this query is used to return the list of presidents that were replaced by \dqlt{Michelle Bachelet}, and the dates they were replaced by her.
Illustrating a join on an edge type, consider the following query:

\begin{dql}
SELECT ?x, ?y
MATCH (?x)-[?e TYPE(?t)]->(?y),
      (?t)=[subproperty of*]=>(parent)
\end{dql}

\noindent This query returns all pairs of nodes with an edge whose type is \dqlt{parent}, or a transitive sub-property of  \dqlt{parent}; for example, Wikidata defines \dqlt{father} and \dqlt{mother} to be sub-properties of \dqlt{parent}.


\paragraph*{Optional matches.} \ql also supports optional graph patterns, which behave akin to left outer joins. For instance, we may wish to query for presidents of Chile, and if available, the president they replaced, and if further available, the president that the latter president replaced. The following \ql query accomplishes this:

\noindent
\begin{dql}
SELECT ?x, ?y, ?z
MATCH (?x)-[?e1 position held]->(President of Chile),
      OPTIONAL {
        (?e1)-[replaces]->(?y)
        OPTIONAL {
          (?y)-[?e2 position held]->(President of Chile),
          (?e2)-[replaces]->(?z)
        }
      }
\end{dql}

\noindent If evaluated on the domain graph of Figure~\ref{fig:dg}, we would get two solutions: $\{ \{ \dqlt{?x} \mapsto \dqlt{Michelle Bachelet}, \dqlt{?y} \mapsto \dqlt{Ricardo Lagos} \},$ $\{ \dqlt{?x} \mapsto \dqlt{Michelle Bachelet}, \dqlt{?y} \mapsto \dqlt{Sebastián Piñera} \} \}$, where in both cases, the variable \dqlt{?z} is left \textit{unbound} as the data is not available for the inner optional graph pattern. As shown, nested optionals are supported, but with the restriction that they form \textit{well-designed patterns} as defined in
\cite{PerezAG09}.

\paragraph*{Limits and ordering.} Some additional operators that MillenniumDB supports are \dqlkw{LIMIT} and \dqlkw{ORDER BY}. These allow us to limit the number of output mappings, and sort the obtained results. For instance, if we need to obtain the ten most recent presidents of Chile, we could accomplish this with the following query:

\begin{dql}
SELECT ?x
MATCH (?x)-[?e position held]->(President of Chile),
      (?e)-[start date]->(?d)
LIMIT 10
ORDER BY DESC (?d)
\end{dql}

\noindent Ordering is always applied before limiting results.

%

\paragraph{Formal definitions.} At~\cite{extended}, we include an extended version of this paper, which contains a full specification of \ql, as well as a formal definition of its semantics. A grammar for generating \ql queries is presented in Appendix A of~\cite{extended}. The abstract syntax and formal semantics of the language is given in Appendix B of~\cite{extended}.




\section{System architecture}
\label{sec:architecture}
In this section, we describe the architecture underlying MillenniumDB. We begin by explaining how one can store \datas and access them from disk. We then outline the process of query evaluation, and also provide some details on less conventional algorithms deployed in this process.

\paragraph*{Storage.} Internally, all objects are represented as 8 byte identifiers. To optimize query execution, identifiers are divided into classes and the first byte of the identifier specifies a class it belongs to. The main classes in a property domain graph $G=(O,\gamma,\lab,\prop)$ are:
\begin{itemize}
	\item \emph{Nodes}, which are objects in the range of $\gamma$. They are divided in two subclasses: \emph{named nodes},  which are objects in the domain graph for which an explicit name is available (e.g. Q320 in Wikidata), and \emph{anonymous nodes}, which are internally generated objects without an explicit name available to the user (similar to blank nodes in RDF~\cite{HoganAMP14}).

	\item \emph{Edges}, which are objects in the domain and range of $\gamma$, and are always anonymous, internally generated objects. 
	
	\item \emph{Values}, which are data objects like strings, integers, etc. These values are classified in two subclasses: \emph{inlined values}, which are values that fit into 7 bytes of the identifier after the mask (e.g. 7 byte strings, integers, etc.), and \emph{external values}, which are values longer than 7 bytes (e.g. long strings).
\end{itemize}
All records stored in MillenniumDB are composed of these identifiers. Therefore, when we write $o\in O$, we are referring to the identifier of the object $o$, with some of the classes above. As we mentioned, the first byte of the 8 byte identifier defines the class, and the remaining 7 bytes are used to store an id (e.g., edges), a name (e.g., named nodes), or data like an integer or a short string (e.g., inlined values). We will later explain how long strings for external values are handled.

To store property \datas, MillenniumDB deploys B+ trees \cite{ramakrishnan00}. For this purpose, we built a B+ tree template for fixed sized records, which store all classes of identifiers. To store a property \data $G=(O,\gamma,\lab,\prop)$, we simply store and index the four components defining it in B+ trees:
\begin{itemize}
	\item \textsc{Objects}(\underline{\textsf{id}}) stores the identifiers of all the objects in the database (i.e., $O$). 
	\item \textsc{DomainGraph}(\textsf{source},\textsf{type},\textsf{target},\underline{\textsf{eid}}) contains all information on edges in the graph (i.e., $\gamma$), where \textsf{eid} is an edge identifier, and \textsf{source}, \textsf{type}, and \textsf{target} can be ids of any class (i.e., node, edge, or value). By default, four permutations of the attributes are indexed in order to aid query evaluation. These are: \textsf{source-target-type-eid}, \textsf{ target-type-source-eid}, \textsf{type-source-target-eid} and \textsf{ type-target-source-eid}. 
	\item \textsc{Labels}(\textsf{object},\textsf{label}) stores object labels (i.e., $\lab$). The value of \textsf{id} can be any identifier, and the values of \textsf{label} are stored as ids. Both permutations are indexed.
	\item \textsc{Properties}(\underline{\textsf{object},\textsf{property}},\textsf{value}) stores the property--value pairs associated with each object (i.e., $\prop$). The \textsf{object} column can contain any id, and \textsf{property} and \textsf{value} are value ids. Aside from indexing the primary key, an additional permutation is added to search objects by property--value pairs.
\end{itemize}

All the B+ trees are created through a bulk-import phase, which loads multiple tuples of sorted data, instead of creating the trees by inserting records one by one. 
%
In order to enable fast lookups by edge identifier, we use the fact that this attribute is the key for the relation. Therefore, we also store a table called \textsc{EdgeTable}, which contains triples of the form $(\textsf{source,\,type,\,target})$, such that the position in the table equals to the identifier of the object $e$ such that $\gamma(e)=(\textsf{source,\,type,\,target})$. This implies that edge identifiers must be assigned consecutive ids starting from zero, and they are generated in this way by MillenniumDB (they are not specified by the user). 
In total, we use ten B+ trees for storing the data. 

To transform external strings and values (longer than 7 bytes) to database object ids and values, we have a single binary file called \textsc{ObjectFile}, which contains all such strings concatenated together. The internal id of an external value is then equal to the position where it is written in the \textsc{ObjectFile}, thus allowing efficient lookups of a value via its id. The identifiers are generated upon loading, and an additional hash table is kept to map a string to its identifier; we use this to ensure that no value is inserted twice, and to transform explicit values given in a query to their internal ids. 
Only (long strings) are currently supported, but the implementation interface allows for adding different value types in a simple manner. 

\paragraph*{Access methods.} All of the stored relations are accessed through linear iterators which provide access to one tuple at a time. All of the data is stored on pages of fixed (but parametrized) size (currently 4kB). The data from disk is loaded into a shared main memory buffer, whose size can be specified upon initializing the MillenniumDB server. The buffer uses the standard clock page replacement policy~\cite{ramakrishnan00}. Additionally, for improved performance, upon initializing the server, it can be specified that the \textit{ObjectFile} be loaded into main memory in order to quickly convert internal identifiers to string and integer values that do not fit into 7 bytes.

\paragraph*{Evaluating a query.} The query execution pipeline follows the standard database template:
\begin{enumerate}
\item The string of the query is parsed into an Abstract Syntax Tree (AST), and checked to be syntactically correct.
\item A logical plan is generated using a depth-first traversal of the AST, instantiating nodes of the tree as logical operators. 
\item Using a visitor pattern, the logical tree is parsed in order to apply some basic simplifications (e.g. pushing constants from filters into the search pattern).
\item Finally, another pass of the logical tree using the visitor-pattern is made in order to create a physical plan.
\end{enumerate}

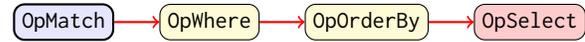
\begin{figure}
\centering
\begin{dql}
SELECT   SelOptions
MATCH    MatchPattern
WHERE    Condition
ORDER BY VarList
\end{dql}

\vspace*{0.2cm}

\begin{tikzpicture}
\node[fill=red!20, draw, rounded corners] (Sel) {$\texttt{OpSelect}$};

\node[left = 0.6cm of Sel, fill=yellow!20, draw, rounded corners] (Fil) {$\texttt{OpOrderBy}$};

\node[left = 0.6cm of Fil, fill=yellow!20, draw, rounded corners] (Root) {$\texttt{OpWhere}$};

\node[left = 0.6cm of Root, fill=blue!10, draw, rounded corners, thick] (Pat) {$\texttt{OpMatch}$};

\draw[<-,thick, color = red] (Sel)--(Fil);
\draw[<-,thick, color = red] (Fil)--(Root);
\draw[<-,thick, color = red] (Root)--(Pat);

\end{tikzpicture}

\caption{A logical plan for a generic MillenniumDB query. Red and blue nodes must always be present in the query plan, while the yellow ones can be omitted. OpMatch is either a single conjunction of atomic patterns (node, edge, or property path patterns), or an \dqlkw{OPTIONAL} pattern.}
\label{fig-logical}
\end{figure}

An illustration of what a logical plan looks like for a generic MillenniumDB query is given in Figure \ref{fig-logical}. In the plan tree, there will be nodes corresponding to clauses such as \dqlkw{MATCH} and \dqlkw{WHERE}. 
We prefix each logical operator with an ``\texttt{Op}" (so the names will be \texttt{OpMatch}, \texttt{OpSelect}, etc.). 
In Figure \ref{fig-logical} each \texttt{OpMatch} is also associated with the node, edge, or property path patterns appearing in it, but we omit those for brevity. 
As we can notice, some parts of the query are mandatory (corresponding to the \texttt{SELECT} and \texttt{MATCH} clauses), while others need not be present.
Finally, a physical plan is generated by assigning a linear iterator returning solution mappings to each \texttt{Op}  node in the logical plan. This allows each \texttt{Op} node to pass its results to the node above following the red arrows in Figure \ref{fig-logical}, thus resulting in a  pipelined evaluation of queries. 

\begin{figure}
  \centering
\begin{dql}
SELECT ?x, ?y
MATCH (?x :human)-[father]->(?y :human)
\end{dql}

\vspace*{0.2cm}  
  
\begin{tikzpicture}
\node[fill=red!20, draw, rounded corners] (Sel) {$\texttt{OpSelect}$};

\node[left = 0.8cm of Sel, fill=blue!10, draw, rounded corners, dashed] (Pat) {
   \alt{ \\
        \texttt{OpLabel(?x, human)}\\
        \texttt{OpLabel(?y, human)}\\
        \texttt{OpConnection(?x, ?\_c0, father, ?y)}}  };
         
\node[fill=blue!20, draw, rounded corners] (le1) at (Pat.north) {$\texttt{OpMatch}$};

\draw[<-,thick, color = red] (Sel)--(Pat);


\end{tikzpicture}
\vspace{3mm}

\resizebox{\columnwidth}{!}{
\includegraphics[scale=1.0]{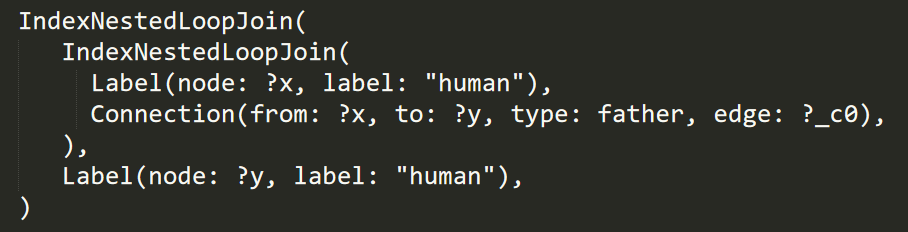}
}

\caption{Generating a physical plan based on a logical plan. Notice that the \texttt{OpMatch} now lists the relations that are to be joined in the query, and an implicit edge variable \texttt{?\_c0} is introduced. The physical plan in this case simply joins the used relations.}
\label{fig-phys}
\end{figure}

The interesting part is evaluating the \texttt{OpMatch} pattern, which is a list of relations that can be edges, labels, properties, or path queries. In essence, evaluating \texttt{OpMatch} is analogous to selecting an appropriate join plan for the relations representing the different elements. This also goes in hand with selecting the appropriate join algorithm for each of the joins. Given that  edges, labels, and properties are all indexed, this will most commonly be index nested loops join. Paths on the other hand are not directly indexed. For this reason, they are currently pushed to the very end of the join plan and joined via a nested-loop join with the remaining part of \texttt{OpMatch} \footnote{We admit that this is not always the best option. However, based on extensive empirical evidence (see Section \ref{sec:benchmark}), this solution seems to be adequate in practice.}. We illustrate a sample query, and a possible plan generated for this query in Figure \ref{fig-phys}.

Currently, MillenniumDB supports two different evaluation mechanisms for evaluating \texttt{OpMatch}:
\begin{itemize}
\item The classical relational optimizer which is based on cost estimation, and tries to order base relations in such a way as to minimize the amount of (intermediate) results. We currently support two modes of execution here:
\begin{itemize}
\item[(i)] Selinger-style join plans \cite{Sciore20} which use dynamic programming to determine the optimal order of relations.
\item[(ii)] In the presence of a large number of relations, a greedy planner \cite{0020812} is used which simply determines the cheapest relation to use in each step.
\end{itemize}
\item A worst-case optimal query plan as described in~\cite{HoganRRS19} is used whenever possible. This approach implements a modified leapfrog algorithm~\cite{Veldhuizen14} in order to minimize the number of intermediate results that are generated.
\end{itemize}

Two particular points of interest are the worst-case optimal query planner for \texttt{OpMatch}, and the way that property paths are evaluated. Both of these deploy state-of-the-art research ideas that are usually not, to the best of our knowledge, implemented in graph database systems. We provide some additional details on these next.

\paragraph*{Worst-case optimal query plan.} Evaluating \texttt{OpMatch} in a worst-case optimal way is done using a modified leapfrog algorithm~\cite{HoganRRS19}. While a classical join plan does a nested for-loop over relations, leapfrog performs a nested for-loop over variables~\cite{Veldhuizen14}. Specifically, the algorithm first selects a variable order for the query, say $(\texttt{?x},\texttt{?y},\texttt{?z})$. It then intersects all relations where the first variable \texttt{?x} appears, and over each solution for \texttt{?x} returned, it intersects all relations where \texttt{?y} appears (replacing \texttt{?x} its current solution), and so on to \texttt{?z}, until all variables are processed and the final solutions are generated. We refer the reader to \cite{Veldhuizen14} and \cite{HoganRRS19} for a detailed explanation.  Two critical aspects for supporting this approach are indices and variable ordering, explained next. 

To support the leapfrog algorithm, we should index all relations in all possible orders of their attributes, which greatly increases disk storage~\cite{HoganRRS19}. In MillenniumDB, we include four orders for \textsc{DomainGraph}, and all orders for \textsc{Labels} and \textsc{Properties}. By considering these orders, we can cover the most common join-types that appear in practice~\cite{BonifatiMT20} by a worst-case optimal query plan. We use the classical relational optimizer if the plan needs an unsupported order or one of the relations uses a path query. 

The leapfrog algorithm requires choosing a variable ordering, which is crucial for its performance. The heuristic we deploy for selecting the variable ordering mixes a greedy approach, and the ideas of the GYO reduction \cite{YuO79}. More precisely, we first order the variables based on the minimal cost of the relations they appear in and resolve ties by selecting the variable that appears in more distinct relations. The variables ``connected" to the first one chosen are then processed in the same manner (where connected meant appearing in the same relation) until the process can not continue. The isolated variables are then treated last.

\paragraph*{Evaluating path query.} For evaluating a path query, the path pattern (2RPQ) is compiled into an automaton, and a ``virtual" cross-product of this automaton and the graph is constructed on-the-fly, and navigated in a breadth-first manner\footnote{BFS is the default evaluation method for property paths. The source code also includes four other algorithms, each with their merits and cons. Detailed exploration of those is outside of the scope of this paper.}, as commonly suggested in the theoretical literature \cite{MendelzonW89,Baeza13,BaierDRV17}. Our assumption is that each path pattern will have at least one of the endpoints assigned before evaluation. This can be done either explicitly in the pattern, or via the remainder of the query. For instance, a property paths pattern \texttt{(Q1)=[P31*]=>(?x)} has the starting point of our search assigned to \texttt{Q1}. On the other hand, \texttt{(?x)=[P31*]=>(?y :Person)} does not have any of the endpoints assigned, however, the \texttt{(?y :Person)} allows us to instantiate \texttt{?y} with any node with the label \texttt{:Person}. 

Intuitively, from a starting node (tagged with the initial state of the automaton), all edges with the type specified by the outgoing transitions from this state are followed. The process is repeated until reaching an end state of the automaton, upon which a result can be returned. This allows a fully pipelined evaluation of path queries, while only requiring at most one pinned page in the buffer (the neighbors of the node on the top of the BFS queue).

Additionally, the BFS algorithm also allows us to return a single shortest path between each pair of endpoints. This can be returned by adding a variable in the property path pattern as follows:
 
\begin{dql}
SELECT *
MATCH (Kevin Bacon)=[?p (actedIn/^actedIn)*]=>(?actor)
\end{dql}
 
\noindent
This query begins at the node corresponding to \texttt{Kevin Bacon}, and looks for all the actors who acted with him in the same movie, or in a movie with someone who acted with him, etc. The variable \texttt{?p} will return a single shortest path between Kevin Bacon and any other actor who can be reached via a path satisfying the query. We note that returning paths comes almost for free, given that they can be reconstructed using the set of visited nodes which is used for bookkeeping in the BFS algorithm.

\section{Benchmarking}
\label{sec:benchmark}
In this section we provide an experimental evaluation of the core features of MillenniumDB. We base our experiments on the Wikidata knowledge graph~\cite{VrandecicK14}, as it is one of the largest and most diverse real-world knowledge graphs that is publicly available, and it provides an extensive query log of real-world queries~\cite{MalyshevKGGB18,BonifatiMT20}. The experiments focus on two fundamental query features: (i) basic graph patterns (BGPs); and (ii) path queries.  We compare the performance of MillenniumDB (MillDB) with several popular persistent graph database engines that support BGPs and at least the Kleene star feature for paths. We publish the data, queries, scripts, and configuration files for each engine online, together with the scripts used to load the data and run the experiments~\cite{benchmark}.

\paragraph*{The engines.} 
We compare the performance of MillenniumDB~\cite{millenniumDB} with five persistent graph query engines. First, we include three popular RDF engines: Jena TDB version 4.1.0~\cite{JenaTDB}, Blazegraph (BlazeG for short) version 2.1.6~\cite{ThompsonPC14}, and Virtuoso version 7.2.6~\cite{Erling12}. We further include a property graph engine: Neo4J community edition 4.3.5~\cite{Webber12}. Finally, we also tested with Jena Leapfrog (Jena LF, for short) -- a version of Jena TDB implementing a worst-case optimal leapfrog algorithm~\cite{HoganRRS19} -- in order to compare the performance of our worst case optimal algorithm with a persistent engine implementing a similar algorithm for evaluating BGPs. We further include an internal baseline, where we test MillenniumDB with (MillDB LF) and without (MillDB NL) the worst-case optimal join algorithm enabled; nested-loop joins are used in the latter case. We remark that MillDB LF is the default version of  MillenniumDB.

\paragraph*{The machine.} All experiments described were run on a single commodity server with an Intel\textregistered Xeon\textregistered Silver 4110 CPU, and 128GB of DDR4/2666MHz RAM, running the Linux Debian 10 operating system with the kernel version 5.10. The hard disk used to store the data was a SEAGATE ST14000NM001G with 14TB of storage. 

\paragraph*{The data.} We use two Wikidata datasets in our experiments. First, in order to compare query performance across various engines, we used the truthy dump version 20210623-truthy-BETA~\cite{WikidataData}, keeping only triples in which (i) the subject position is a Wikidata entity, and (ii) the predicate is a direct property. We call this dataset \textit{Wikidata Truthy}. The size of the dataset after this process was 1,257,169,959 triples. The simplification of the dataset is done to facilitate comparison across multiple engines, specifically to keep data loading times across all engines manageable while keeping the nodes and edges necessary for testing the performance of BGPs and property paths. However, MillenniumDB was designed with the complete version of Wikidata -- including qualifiers, references, etc. -- in mind. We thus specifically test MillenniumDB on a second dataset, called \textit{Wikidata Complete}, which comprises the entire Wikidata knowledge graph, including the aforementioned features. The version JSON dump 0201102-all.json was used, preprocessed and mapped to our data model. More details about the preparation of the datasets, as well as the datasets themselves, can be found online~\cite{benchmark}.


\begin{table}[t]
\caption{The dataset size when loaded into each engine. The original RDF dataset consists of roughly 1.25 billion triples. \label{tab-data}}
\footnotesize
\begin{tabular}{cccccc}
\toprule
MillDB & BlazeG & Jena & Jena LF & Virtuoso & Neo4J \\
\midrule
203GB & 70GB & 110GB & 195GB &  70GB & 112GB \\
\bottomrule
\end{tabular}

\end{table}

The size of the \textit{Wikidata Truthy} dataset, when loaded into the respective systems, is summarized in Table \ref{tab-data}. To store the data, default indices were used on Jena TDB, Blazegraph and Virtuoso. Jena LF stores three additional permutations of the stored triples to efficiently support the leapfrog algorithm for any join query. Neo4j by default creates an index for edge types (as of version 4.3.5). To support faster searches for particular entities and properties, we also created an index linking a Wikidata identifier (such as, e.g., Q510) to its internal id in Neo4j. We also tried to index literal values in Neo4j, but the process failed (the literals are still stored). MillenniumDB uses extra disk space, because of the additional indices needed to support the domain graph model. Similarly, Jena LF indexes more permutations than Jena TDB in order to support worst-case optimal joins, hence using more space.

For scale-up experiments, we subsequently loaded \textit{Wikidata Complete} into MillenniumDB, which further exploits the domain graph model.\footnote{It is important to note that JSON and RDF dumps of Wikidata do not result in precisely the same knowledge graph due to some restrictions of the particular reification used in RDF; however, they do result in very similar knowledge graphs.} In this version, we model qualifiers (i.e. edges on edges), put labels on objects, and assign them properties with values. We use properties to store the language value of each string in Wikidata, and also to model elements of complex data values (e.g., for coordinates we would have objects with properties latitude and longitude, and similarly for amounts, date/time, limits, etc.). Each object representing a complex data value also has a label specifying its data type (e.g. \texttt{coord} for geographical coordinates). Qualifiers were loaded in their totality (i.e., not only the preferred qualifiers, but all specified ones). The only elements excluded from the data dump (for now) were sitelinks and references. This full version of Wikidata resulted in a knowledge graph with roughly 300 million objects, participating in $4.3 \times 10^9$ edges. The total size on disk of this data was 827GB in MillenniumDB, i.e., more than four times larger than \textit{Wikidata Truthy}. In Section~\ref{ss-full}, we illustrate that this data bloat does not significantly affect query evaluation, and show comparable times to the ones MillenniumDB achieves on the smaller \textit{Wikidata Truthy} dataset. 

\paragraph*{How we ran the queries.} We detail the query sets used for the experiments in their respective subsections. 
To simulate a realistic database load, we do not split queries into cold/hot run segments. Rather, we run them in succession, one after another, after a cold start of each system (and after cleaning the OS cache). 
This simulates the fact that query performance can vary significantly based on the state of the system buffer, or even on the state of the hard drive, 
or the state of OS's virtual memory. For each system, queries were run in the same order. 
We record the execution time of each individual query, which includes iterating over all results. We set a limit of 100{,}000 distinct results for each query, again in order to enable comparability as some engines showed instability when returning larger results. Jena and Blazegraph were assigned 64GB or RAM, and Virtuoso was set up with 64GB or more of RAM as is recommended. Neo4J was run with default settings, while MillenniumDB had access to 32GB for main-memory buffer, and it uses an additional 10GB for in-memory dictionaries.

\paragraph*{Handling timeouts.} We defined a timeout of 10 minutes per query for each system. As we will see, MillenniumDB was the only system not to time out on any query. Apart from that, we note that most systems had to be restarted upon a timeout as they often showed instability, particularly while evaluating path queries. This was done without cleaning the OS cache in order to preserve some of the virtual memory mapping that the OS built up to that point. In comparison, even when provided with a very small timeout window, MillenniumDB managed to return a non-trivial amount of query results on each query, and did not need to be restarted, thus allowing us to handle timeouts gracefully. 

%

\subsection{Basic Graph Patterns}
\label{ss-bgps}
We focus first on basic graph pattern queries. To test different query execution strategies of MillenniumDB, we use two benchmarks: \emph{Troublesome BGPs} and \emph{Complex BGPs}, which are described next.

\paragraph*{Troublesome BGPs} The Wikidata SPARQL query log contains millions of queries~\cite{MalyshevKGGB18}, but many of them are trivial to evaluate. We thus decided to generate our benchmark from more challenging cases, i.e., a smaller log of queries that timed-out on the Wikidata public endpoint~\cite{MalyshevKGGB18}. From these queries we extracted their BGPs, removing duplicates (modulo isomorphism on query variables). We further removed queries that give zero results over \textit{Wikidata Truthy}. We distinguish queries consisting of a single triple pattern (\emph{Single}) from those containing more than one triple pattern (\emph{Multiple}). The former set tests the triple matching capabilities of the systems, whereas queries in the latter set test join performance. 
\textit{Single} contains 399 queries, whereas \textit{Multiple} has 436 queries. 

\begin{figure}[t]
\centering
\hbox{
\includegraphics[width=0.339\columnwidth]{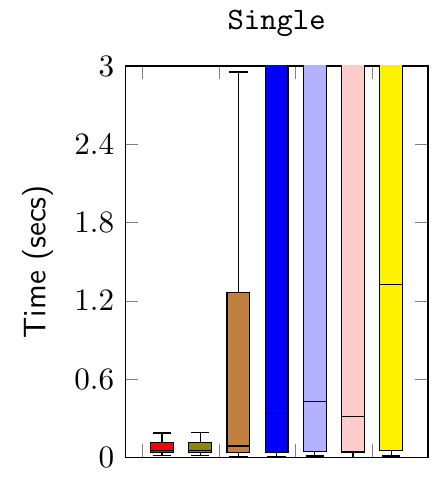}
\includegraphics[width=0.320875\columnwidth]{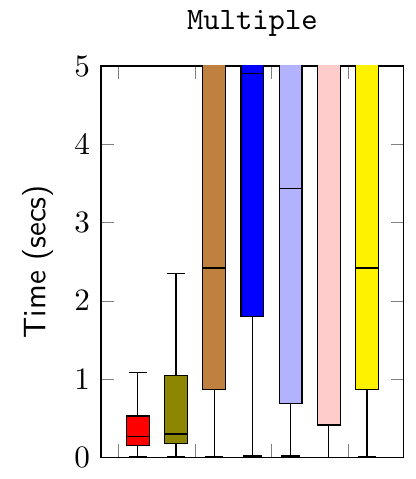}
\includegraphics[width=0.339\columnwidth]{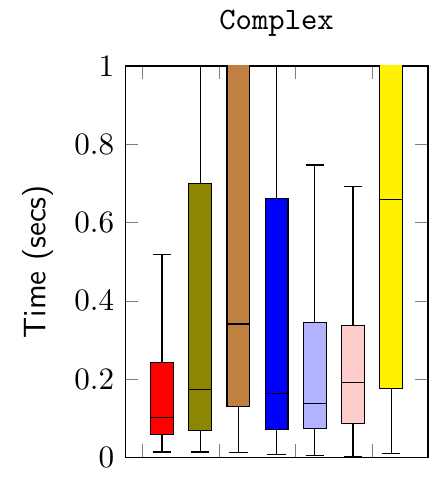}
}

    \vspace{0.3cm}
    \begin{tikzpicture}
       \begin{customlegend}[legend columns=4,legend style={draw=none,column sep=1ex, font=\footnotesize, cells={anchor=west}, at={(0.5,-0.15)},
      anchor=north, /tikz/every even column/.append style={column sep=0.4cm}},legend entries={\small\textsf{MillDB LF}, \small\textsf{MillDB NL}, \small\textsf{Blazegraph}, \small\textsf{Jena}, \small\textsf{Jena LF}, \small\textsf{Virtuoso}, \small\textsf{Neo4J}}]
          \addlegendimage{black,fill=red, ybar stacked}
          \addlegendimage{black,fill=olive, ybar stacked}          
          \addlegendimage{black,fill=brown,ybar stacked}
          \addlegendimage{black,fill=blue,ybar stacked}
          \addlegendimage{black,fill=blue!30,ybar stacked}
          \addlegendimage{black,fill=red!20,ybar stacked}
	  \addlegendimage{black,fill=yellow,ybar stacked}
       \end{customlegend}
    \end{tikzpicture}
\caption{Boxplots for the query time distribution on the set of troublesome Single BPG queries (left), troublesome Multiple BPG queries (middle), and Complex BGP queries (right).}
\label{fig:bgp-queries1}
\end{figure}

Table~\ref{tab:queries1} (top and middle) summarizes the query times on this set, whereas Figure~\ref{fig:bgp-queries1} (left and middle) shows boxplots with more detailed statistics on the distributions of runtimes.

\begin{table}
\caption{Summary of query times, in seconds, for BGPs. Average and median are for all the queries, including timeouts. \label{tab:queries1}}
\footnotesize%
\setlength{\tabcolsep}{3.2pt}%
\begin{tabular}{crrrrrrr}
\toprule
\multicolumn{8}{c}{\textit{Single}} \\
\midrule
		  & MillDB LF & MillDB NL & BlazeG & Jena & Jena LF & Virtuoso & Neo4J \\
\midrule
Supported         & 399  & 399  & 399  & 399   & 395   & 399   & 394\\
Error             & 0    & 0    & 0    & 0     & 4     & 0     & 5\\
Timeouts          & 0    & 0    & 0    & 0     & 0     & 0     & 0\\
Average           & 0.07 & 0.07 & 2.21 & 14.10 & 10.08 & 2.22  & 28.00\\
Median            & 0.05 & 0.05 & 0.09 & 0.34  & 0.44  & 0.32  & 1.33\\
\midrule
\end{tabular}

\vspace{2mm}

\begin{tabular}{crrrrrrr}
\toprule
\multicolumn{8}{c}{\textit{Multiple}} \\
\midrule
		  & MillDB LF & MillDB NL & BlazeG & Jena & Jena LF & Virtuoso & Neo4J\\
\midrule
Supported         & 436   & 436   & 436   & 426   & 418   & 436   & 405 \\
Error             & 0     & 0     & 0     & 10    & 18    & 0     & 31\\
Timeouts          & 0     & 1     & 3     & 0     & 0     & 0     & 0\\
Average           & 4.77 & 11.01 & 31.79 & 35.43 & 16.78 & 7.87  & 75.55\\
Median            & 0.27  & 0.30  & 2.42  & 4.90  & 3.39  & 5.11  & 6.84\\
\hline
\end{tabular}
\vspace{2mm}

\begin{tabular}{lrrrrrrr}
\toprule
\multicolumn{8}{c}{\textit{Complex}} \\
\midrule
		  & MillDB LF & MillDB NL & BlazeG & Jena & Jena LF & Virtuoso & Neo4J\\
\midrule
Supported         & 850   & 850   & 850   & 850   & 850   & 850   & 850 \\
Error             & 0     & 0     & 0     & 2    & 0    & 0     & 10\\
Timeouts          & 0     & 1     & 2     & 0     & 0     & 0     & 0\\
Average           & 0.37 & 3.36 & 4.63 & 3.37 & 0.88 & 1  & 17.92\\
Median            & 0.1  & 0.17  & 0.34  & 0.16  & 0.14  & 0.19  & 0.66\\
\midrule
\end{tabular}
\end{table}
MillenniumDB sharply outperforms the alternatives on the \textit{Single} queries.
The next best performers on average, Blazegraph
and Virtuoso, are more than 30 times slower for average runtimes. In terms of medians, Blazegraph comes closest, but is still almost twice as slow as 
MillenniumDB. It is also much less stable: the third quartile of Blazegraph is 
an order of magnitude higher than the most costly queries in MillenniumDB.%
\footnote{Excluding the outliers, which are not included between whiskers.} The
other systems tested are slower on \textit{Single}, with their averages being two orders of magnitude
higher than those of MillenniumDB and their medians being higher than all of the 
times shown for MillenniumDB.

MillenniumDB also outperforms the other systems on queries of the \textit{Multiple} set. Its medians are an
order of magnitude faster than those of Blazegraph, the next best contender.
Indeed, the median of Blazegraph is an order of magnitude higher than all the 
queries in MillenniumDB (excluding outliers). Even the first quartile of 
Blazegraph is close to the most costly queries in MillDB LF. The difference is
less sharp for averages, but still MillDB LF takes 60\% of the time of Virtuoso,
the next best contender. The comparison between both query strategies of 
MillenniumDB shows that, though both medians are close, the worst-case-optimal 
MillDB LF offers more stable times than the nested-loop strategy of MillDB NL.

\paragraph{Complex BGPs.}
This is a benchmark used to test the performance of worst-case optimal joins~\cite{HoganRRS19}. Here, 17 different complex join patterns were selected, and 50 different queries generated for each pattern, resulting in a total of 850 queries. Figure~\ref{fig:bgp-queries1} (right), and Table~\ref{tab:queries1} (bottom), show the resulting query times. In this case, the difference between the two evaluation strategies of MillenniumDB is more clear. The worst-case-optimal version (MillDB LF) is not only considerably more stable than the nested-loop version (MillDB NL), but also twice as fast in the median. After MillDB LF, the next-best competitor is Jena LF, showing the benefits of worse-case optimal joins. Virtuoso follows not far behind, while MillDB NL, Jena, Blazegraph and Neo4j are considerably slower. Overall, MillDB LF offers the 
best performance for every statistic shown in the plot.


%
%

\subsection{Property Paths}
\label{ss-paths}

\begin{figure}[t]
\includegraphics[width=0.45\columnwidth]{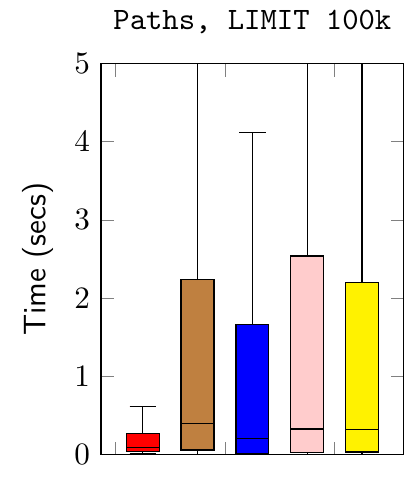}

    \vspace{0.3cm}
    \begin{tikzpicture}
       \begin{customlegend}[legend columns=5,legend style={draw=none,column sep=1ex, font=\footnotesize, cells={anchor=west}, at={(0.5,-0.15)},
      anchor=north, /tikz/every even column/.append style={column sep=0.2cm}},legend entries={\small\textsf{MillDB}, \small\textsf{Blazegraph}, \small\textsf{Jena}, \small\textsf{Virtuoso}, \small\textsf{Neo4J}}]
          \addlegendimage{black,fill=red, ybar stacked}
          \addlegendimage{black,fill=brown,ybar stacked}
          \addlegendimage{black,fill=blue,ybar stacked}
          \addlegendimage{black,fill=red!20,ybar stacked}
          \addlegendimage{black,fill=yellow,ybar stacked}          
       \end{customlegend}
    \end{tikzpicture}
\caption{Boxplots of query times on property paths, limiting the results to 100{,}000.}
\label{fig:rpqs}
\end{figure}

To test the performance of path queries, we extracted 2RPQ expressions from a log of queries that timed out on the Wikidata endpoint~\cite{MalyshevKGGB18}. 
The original log has 2110 queries. After removing queries that yield no results (due to missing data), we ended up with 1683 queries. These were run in succession, each restricted to return at most 100{,}000 results.
Each system was started after cleaning the system cache, and with a timeout of 10 minutes. Since these are originally SPARQL queries, not all of them were supported by Neo4J given the restricted regular-expression syntax it supports. We remark that MillenniumDB and Neo4J were the only systems able to handle timeouts without being restarted.\footnote{In fact, MillenniumDB did not give any timeouts. However, we re-ran the experiments with a lower timeout, and observed that the system could recover from interrupting the query gracefully and was able to return the results found before being interrupted.} In this comparison we did not include MillDB NL, nor Jena LF, since these deploy precisely the same execution strategy for property paths as do MillDB LF, and Jena, respectively. The experimental results are summarized in Table~\ref{tab-pp1} and Figure~\ref{fig:rpqs}.

\begin{table}
\caption{Summary of query times, in seconds, for property paths, when limiting the queries to 100{,}000 results. Average and median are for all the queries, including errors and timeouts. Avg. succ. and median succ. excludes the queries that gave a timeout or an error for that particular engine.}
\label{tab-pp1}
\footnotesize%
\begin{tabular}{lrrrrr}
\toprule
		& MillDB & BlazeG & Jena 	& Virtuoso & Neo4J \\
\midrule
Supported 	& 1683 	& 1683	& 1683 	& 1683	& 1622 \\
Error 		& 0	& 2	& 19 	& 55	& 42 \\
Timeouts    & 0	& 44	& 41 	& 4	& 42 \\
Average 	& 1.1	& 27.6	& 22.8 	& 5.8 	& 23.3 \\
Median 		& 0.094	& 0.396	& 0.207 & 0.325 & 0.328 \\
Avg. succ. 	& 1.1	& 11.2	& 6.3 	& 1.8 	& 8.0 \\
Median succ. 	& 0.094	& 0.382	& 0.175 & 0.294 & 0.277 \\
\bottomrule
\end{tabular}
\end{table}

%

We can observe that, again, MillenniumDB is the fastest and has the most stable performance, being able to run all the queries. Its average is near a second, i.e., five times faster than the next best contender (Virtuoso). Its median, below 0.1 seconds, is half the next one (Jena's). Even after removing the queries that timed-out on the other systems, they are considerably slower than MillenniumDB. 
In particular, if we only consider the queries that run successfully on Virtuoso (i.e., excluding the 59 queries that timed-out or gave an error), we get an average time of 0.85 seconds and a median time of 0.086 seconds on MillenniumDB, less than half the times of Virtuoso. The boxplots further show the 
stability of MillenniumDB: 
the medians of
other baselines are over the third quartile of MillenniumDB. 
Their third quartile is 5--10 times higher than that of 
MillenniumDB, and higher than its topmost whisker. 

To further test robustness,
we also ran all of the queries {\it without limiting the output size} on MillenniumDB. In this test, the engine timed out in only 15 queries, each returning between 800 thousand and 44 million results before timing out. When running queries to completion, MillenniumDB averaged 13.4 seconds per query (8 seconds if timeouts were excluded), with a mean of 0.1 seconds (both with and without timeouts).

\subsection{Wikidata Complete} 
\label{ss-full}
To show the scalability of MillenniumDB, and to leverage its domain graph, we ran experiments with \textit{Wikidata Complete} to measure query performance. We ran the same queries from the four benchmarks above (\textit{Single}, \textit{Multiple}, and \textit{Complex} BGPs, as well as property paths). The number of outputs on the two versions of the data, while not the same, was within the same order of magnitude averaged over all the queries. The results are presented in Table \ref{tab-full}. As we can observe, MillenniumDB shows no deterioration in performance when a larger database is considered for similar queries. This is mostly due to the fact that the buffer only loads the necessary pages into the main memory, and will probably require a rather similar effort in both cases. We also note that, again, no queries resulted in a timeout over the larger dataset.

\begin{table}
\caption{Average and median runtimes, in seconds, for MillenniumDB on the complete version of Wikidata with the same query load as in the previous subsections.}
\label{tab-full}
\footnotesize
\begin{tabular}{lrrrr}
\toprule
		& Single & Multiple & Complex & Paths \\
\midrule
Average 	& 0.08	& 4.04	& 0.35 	& 1.04  \\
Median 		& 0.07	& 0.28	& 0.095 &  0.10  \\ 
\bottomrule
\end{tabular}
\end{table}

\section{Conclusions and looking ahead}
\label{sec:concl}
This paper presents MillenniumDB: a persistent, open-source, graph database engine, which implements the \data data model.

Domain graphs adopt the natural idea of adding edge ids to directed labeled edges in order to concisely model higher-arity relations in graphs, as needed in Wikidata, without the need for reserved vocabulary or reification. They can naturally represent popular graph models, such as RDF and property graphs, and allow for combining the features of both models in a novel way. While the idea of using edge ids as a hook for modeling higher-arity relations in graphs is far from new (see, e.g.,~\cite{HernandezHK15,IlievskiGCDYRLL20,LassilaSBBBKKLST}), it is an idea that is garnering increased attention as a more flexible and concise alternative to reification.
To the best of our knowledge, our work is the first to propose a formal data model that incorporates edge ids, a query language that can take advantage of them, and a fully-fledged graph database engine that supports them by design. We also propose to optionally allow (external) annotations on top of the graph structure, thus facilitating better compatibility with property graphs, whereby labels and property--values can be added to graph objects without adding new nodes and edges to the graph~itself.

With respect to the query language, we have proposed a new query syntax inspired by Cypher, but that additionally enables users to take full advantage of the domain graph model by (optionally) referencing edge ids in their queries, and performing joins on any element of the domain graph. We further combine useful features present in both Cypher and SPARQL, in order to provide additional expressivity, such as returning the shortest path witnessing a result for a path query (as captured by a 2RPQ expression). 

In the implementation of MillenniumDB, we combine both tried-and-trusted techniques that have been successfully used in relational database pipelines for decades~\cite{ramakrishnan00} (e.g., B+ trees, buffer managers, etc.), with promising state-of-the-art algorithms for computing worst case optimal joins (leapfrog~\cite{Veldhuizen14}) and evaluating path queries (guided by an an automaton \cite{MendelzonW89,Baeza13}). Our experiments over Wikidata, considering real-world queries and data at large-scale, show that this combination outperforms other persistent graph database engines that are commonly found in practice.


Looking to the future, we foresee extensions such as: returning entire graphs, supporting more complex path constraints, returning sets of paths, path algebra, just to name a few. Regarding more practical features, we aim to add support for full transactions, compact index structures, keyword search, a graph update language, existing graph query languages, and more besides. More importantly, given that MillenniumDB is published as an open source engine, we hope that the research community can view the MillenniumDB code base as a sort of a sandbox for incorporating their novel algorithms and ideas into a modern graph database, without the need to remake storage, indexing, access methods, or query parsers. We also wish to explore the deployment of MillenniumDB for key use-cases; for example, we plan to provide and host an alternative query service for Wikidata, which may help to prioritize the addition of novel features and optimizations as needed in practice.




\begin{acks}
This work was supported by ANID -- Millennium Science Initiative Program -- Code ICN17\_002.
\end{acks}

\bibliographystyle{ACM-Reference-Format}

\newpage
\bibliography{nourlbiblio}


\newpage 
\appendix

\onecolumn
\section*{Appendix}

\section{Syntax of MillenniumDB queries}
\label{sec-app-syntax}

The language supported by MillenniumDB borrows syntax from popular graph query languages SPARQL \cite{HarrisS13} and Cypher \cite{FrancisGGLLMPRS18}. General structure of queries in MillenniumDB can be visualized as follows:

\begin{figure}[H]
\begin{center}
\begin{dql}
SELECT Selectors
MATCH  MatchPattern
WHERE  Condition
ORDER BY OrderSelectors 
LIMIT Number
\end{dql}
\end{center}
\caption{General structure of queries in MillenniumDB}
\label{fig-gen}
\end{figure}

Intuitively, the \textsf{SELECT} clause specifies which of the matched variables will be returned, while the \textsf{MATCH} clause specifies the basic or navigational graph pattern which we will look for in our graph. The \textsf{WHERE} clause is used to filter result based on a selection, usually by restricting the values of some of the attributes of a matched object. The \textsf{ORDER BY} allows us to reorder the results based on the values of some output variables, while \textsf{LIMIT} cuts off the evaluation after a specific number of results had been found. 

\begin{figure*}
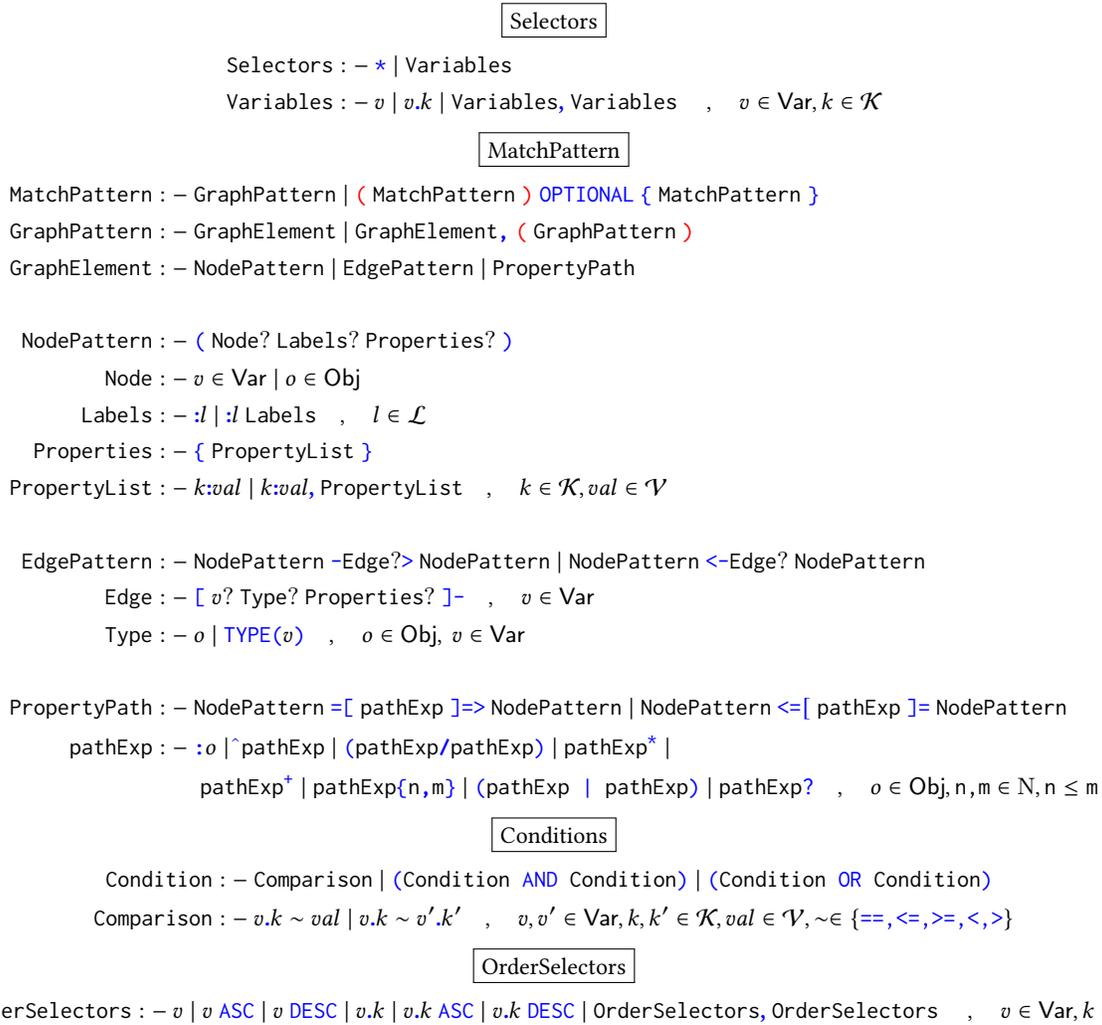

\begin{center}
\fbox{Selectors}
\end{center}
\begin{align*} 
\texttt{Selectors} & :- \ \textcolor{blue}{\texttt{*}} \ | \ \texttt{Variables}\\
\texttt{Variables} & :- \ v \ | \ v\textcolor{blue}{\textbf{.}}k \ | \ \texttt{Variables} \textcolor{blue}{\textbf{,}}\ \texttt{Variables}
\ \quad , \quad v\in \vars, k \in \keys
\end{align*}

\begin{center}
\fbox{MatchPattern}
\end{center}
\begin{align*} 
\texttt{MatchPattern} & :- \ \texttt{GraphPattern}\ |\   \textcolor{red}{\texttt{(}}\ \texttt{MatchPattern}\ \textcolor{red}{\texttt{)}} \ \textcolor{blue}{\texttt{OPTIONAL}} \ \textcolor{blue}{\texttt{\{}} \ \texttt{MatchPattern} \ \textcolor{blue}{\texttt{\}}}\\ 
\texttt{GraphPattern} & :- \ \texttt{GraphElement} \ | \  \texttt{GraphElement} \textcolor{blue}{\texttt{\textbf{,}}} \  \textcolor{red}{\texttt{(}} \ \texttt{GraphPattern} \ \textcolor{red}{\texttt{)}}\\
\texttt{GraphElement} & :- \ \texttt{NodePattern} \ |\ \texttt{EdgePattern} \ |\ \texttt{PropertyPath}\\
 & \\
\texttt{NodePattern} & :- \ \textcolor{blue}{\texttt{(}} \ \texttt{Node}? \ \texttt{Labels}? \ \texttt{Properties}? \ \textcolor{blue}{\texttt{)}}\\
\texttt{Node} & :- \ v \in \vars \ | \ o\in \objs\\
\texttt{Labels} & :- \ \textcolor{blue}{\textbf{:}}l \ | \ \textcolor{blue}{\textbf{:}}l \ \texttt{Labels} \quad , \quad l\in \labels\\
\texttt{Properties} & :- \ \textcolor{blue}{\texttt{\{}} \ \texttt{PropertyList} \ \textcolor{blue}{\texttt{\}}}\\
\texttt{PropertyList} & :- \ k\textcolor{blue}{\textbf{:}}val \ | \ k\textcolor{blue}{\textbf{:}}val\textcolor{blue}{\textbf{,}}\ \texttt{PropertyList}  \quad , \quad k\in \keys, val\in \values \\
 & \\
 \texttt{EdgePattern} & :- \ \texttt{NodePattern} \ \texttt{\textcolor{blue}{-}} \texttt{Edge}? \texttt{\textcolor{blue}{>}} \ \texttt{NodePattern} \ |\            \texttt{NodePattern} \ \texttt{\textcolor{blue}{<-}} \texttt{Edge}? \texttt{\textcolor{blue}{}} \ \texttt{NodePattern} \\
\texttt{Edge} & :- \ \texttt{\textcolor{blue}{[}} \ v? \ \texttt{Type}? \ \texttt{Properties}? \ \texttt{\textcolor{blue}{]-}} \quad , \quad v\in \vars \\
 \texttt{Type} & :-\ o \ | \ \textcolor{blue}{\texttt{TYPE(}}v\textcolor{blue}{\texttt{)}} \quad , \quad o\in \objs, \ v\in \vars \\
  &  \\
 \texttt{PropertyPath} & :- \ \texttt{NodePattern} \ \textcolor{blue}{\texttt{=[}} \ \texttt{pathExp} \ \textcolor{blue}{\texttt{]=>}} \ \texttt{NodePattern} \ |\  \texttt{NodePattern} \ \textcolor{blue}{\texttt{<=}} \textcolor{blue}{[} \ \texttt{pathExp} \ \textcolor{blue}{\texttt{]=}} \ \texttt{NodePattern}\\
 \texttt{pathExp} & :- \ \texttt{\textcolor{blue}{\textbf{:}}}o  \ |\ \textcolor{blue}{\hat{ }} \ \texttt{pathExp} \ |\  \textcolor{blue}{\texttt{(}}\texttt{pathExp\textcolor{blue}{\textbf{/}}pathExp}\textcolor{blue}{\texttt{)}} \ |\ \texttt{pathExp}^{\textcolor{blue}{\texttt{*}}} \ |\\
 & \quad \quad \texttt{pathExp}^{\textcolor{blue}{\texttt{+}}} \ | \  \texttt{pathExp\textcolor{blue}{\{}n\textcolor{blue}{\textbf{,}}m\textcolor{blue}{\}}} \ |\ \textcolor{blue}{\texttt{(}}\texttt{pathExp \textcolor{blue}{|} pathExp}\textcolor{blue}{\texttt{)}} \ |\ \texttt{pathExp}\textcolor{blue}{\texttt{?}} \quad , \quad o\in \objs, \texttt{n,m}\in \mathbf{N}, \texttt{n}\leq \texttt{m}
\end{align*}

\begin{center}
\fbox{Conditions}
\end{center}
\begin{align*} 
\texttt{Condition} & :- \ \texttt{Comparison} \ | \ \textcolor{blue}{\texttt{(}}\texttt{Condition \textcolor{blue}{AND} Condition}\textcolor{blue}{\texttt{)}} \ | \  \textcolor{blue}{\texttt{(}}\texttt{Condition \textcolor{blue}{OR} Condition}\textcolor{blue}{\texttt{)}}\\
\texttt{Comparison} & :- \ v\textcolor{blue}{\textbf{.}}k \sim val \ | \ v\textcolor{blue}{\textbf{.}}k \sim v'\textcolor{blue}{\textbf{.}}k' \quad , \quad v, v'\in \vars, k, k'\in \keys, val \in \values, \sim \in \{\textcolor{blue}{\texttt{==,<=,>=,<,>}}\}
\end{align*}

\begin{center}
\fbox{OrderSelectors}
\end{center}
\begin{align*} 
\texttt{OrderSelectors} & :- \ v \ | \ v \ \textcolor{blue}{\texttt{ASC}} \ | \ v \ \textcolor{blue}{\texttt{DESC}} \ | \ v\textcolor{blue}{\textbf{.}}k \ | \ v\textcolor{blue}{\textbf{.}}k \ \textcolor{blue}{\texttt{ASC}} \ | \ v\textcolor{blue}{\textbf{.}}k \ \textcolor{blue}{\texttt{DESC}} \ | \ \texttt{OrderSelectors}\textcolor{blue}{\textbf{,}}\ \texttt{OrderSelectors} \ \quad , \quad v\in \vars, k \in \keys
\end{align*}

\caption{Specification of query patterns in MilleniumDB. Blue symbols are taken literally. Red brackets in \texttt{MatchPattern} and \texttt{GraphPattern} are grouping specification for unambiguous parsing, and are not specified when writing the query. The \texttt{Number} going in the \texttt{LIMIT} command is any integer.}
\label{fig-syntax}
\end{figure*}

We define the formal syntax MilleniumDB query language in Figure \ref{fig-syntax}.
As an example of a query occupying many of these elements, consider the following:

\begin{dql}
SELECT ?x.age, ?y.age, ?z
MATCH (?x :Person)-[?e knows]->(?y :Person {country:"Chile"})
      OPTIONAL {(?x)-[studiedAt]->(?z)}
WHERE ?x.age >= 35 AND ?e.since == "1/1/2020"
ORDER BY ?x.age ASC, ?y.age DESC
LIMIT 1000
\end{dql}

\noindent When evaluated on a social network graph, the query is trying to find the ages of all pairs of people such the first knows the second. Additionally, it is stating that the second person lives in Chile, that the first one has age greater than or equal 35, and that the edge connecting them was created on \textsf{1/1/2020}. If the university where the first person studied is known, this is also returned (and is otherwise \texttt{null}). The results are ordered such that the age of the first person is ascending, and the age of the second person is descending. Finally, only 1000 results are solicited. The syntax here can be build rater easily using the specification of Figure \ref{fig-syntax}. The most interesting part of the query is in the \texttt{MatchPattern}, which is composed of an \texttt{EdgePattern},  \texttt{(?x :Person)-[?e knows]->(?y :Person \{country:"Chile"\})},  and an optional pattern \texttt{(?x)-[studiedAt]->(?z)}, which is itself an \texttt{EdgePattern}.

\section{Formal Definition of Domain Graph Queries}
\label{sec-app-semantics}

Queries in MillenniumDB are
based on an abstract notion of \textit{domain graph query}, which
generalise the types of graph patterns used
by modern graph query languages~\cite{AnglesABHRV17}. This query
abstraction provides modularity in terms of how the database is
constructed, flexibility in terms of what concrete query syntax is
supported, and allows for defining its semantics and studying its
theoretical properties in a clean way.

This section provides the formal definition of the
MillenniumDB query language.
From now on, assume an infinite
set $\vars$ of variables disjoint with the set of objects $\objs$.

\subsection{Basic graph patterns}

At the core of domain queries are basic graph patterns.\footnote{Basic graph patterns correspond to conjunctive queries (CQs) over graphs.} 
A \emph{basic graph pattern} is defined as a pair $(V,\phi)$ such that $\phi : (\objs \cup \vars) \rightarrow (\objs \cup \vars) \times (\objs \cup \vars) \times (\objs \cup \vars)$ is a partial mapping with a finite domain and $V \subseteq \Var{\phi}$, where $\Var{\phi}$ is the set of variables occurring in the domain or in the range of $\phi$. Thus, $\phi$ can be thought as a domain graph that allows a variable in any position, together with a set $V$ of output variables (hence the restriction that each variable in $V$ occurs in $\phi$).

The evaluation of a basic graph pattern returns a set of solution
mappings.  A \textit{solution mapping} (or simply \textit{mapping}) is
a partial function $\map: \vars \rightarrow \objs$. 
The \emph{domain} of a mapping $\map$, denoted by $\Dom{\map}$, is the set of variables on which $\map$ is defined. 
Given $v \in \vars$ and $o \in \objs$, we use $\mu(v) = o$ to denote that $\mu$ maps variable $v$ to object $o$. 
Besides, given a set $V'$ of variables, the term $\mu_{|_{V'}}$ is used to denote the mapping obtained by restricting $\mu$ to $V'$, that is, $\mu_{|_{V'}} :
(\Dom{\mu} \cap V') \to \objs$ such that $\mu_{|_{V'}}(v) = \mu(v)$ for every $v \in (\Dom{\mu} \cap V')$ (notice that $V'$ is not necessarily a subset of $\Dom{\mu}$).
Finally, for the sake of presentation, we assume that $\mu(o)=o$, for
all $o\in \objs$.

The evaluation of a basic graph pattern $B = (V,\phi)$ over a property-domain graph 
$G=(O,\gamma)$, denoted by $\semp{B}{G}$, is defined as
  $\semp{B}{G} = \{ \map_{|_V} \mid \mu \in \semp{\phi}{G}\}$,
where:
\begin{eqnarray*}
  \semp{\phi}{G} \ = \ \{ \map \mid
  \Dom{\map} = \Var{\phi} \text{ and if } \phi(a) = (a_1,a_2,a_3),
  \text{then } \gamma(\map(a)) = ( \map(a_1), \map(a_2), \map(a_3)  ) \}.
\end{eqnarray*}
For example, consider the basic graph pattern $(V,\phi)$ where $V = \{ v_2, v_4, v_6\}$ and $\phi$ is given by the assignments:
\begin{verse} 
$\phi(v_1) = (v_2, \texttt{position held}, \texttt{President of Chile})$,\\
$\phi(v_3) = (v_1, \texttt{start date}, v_4)$, and\\
$\phi(v_5) = (v_1, \texttt{replaces}, v_6)$.
\end{verse}
In Figure \ref{fig:bgp}, we provide a graphical representation of the above graph pattern, and the solution mappings obtained by evaluating the graph pattern over the property domain graph shown in Figure~\ref{fig:dg}. The solution mappings are presented as a table with columns $v_2$, $v_4$, $v_6$ (i.e. the variables in $V$), and each row represents an individual mapping. In our definitions, different variables may map to the same object in a single solution. Thus, our notion of evaluation follows a homomorphism-based semantics, similar to query languages such as SPARQL~\cite{AnglesABHRV17}.\footnote{Isomorphism-based semantics~\cite{AnglesABHRV17} -- such as Cypher's no-repeated-edge semantics, which disallows the same solution to use the same edge twice -- can be emulated by filtering solutions after they are generated.}

\begin{figure}[tb]
\setlength{\vgap}{1.2cm}
\setlength{\hgap}{4cm}
\centering

\begin{tabular}[t]{ll}
\begin{tabular}{l}
\begin{tikzpicture}

\node[iri,anchor=center,minimum width=7.5cm,minimum height=0.7cm,dashed] (e1) {};

\node[anchor=south east,above left=0.1ex of e1.south east] (e1i) {$v_1$};

\node[iri,right=1ex of e1.west,anchor=west] (sp1) {$v_2$};
\node[iri,right=\hgap of sp1] (ch1) {President of Chile}
  edge[arrin] node[iri,draw=none] {position held} (sp1);
  
\node[iri,below=\vgap of sp1] (v4) {$v_4$}
  edge[arrin] node[lab] {start date} (e1);

\node[iri,below=\vgap of ch1] (mb) {$v_6$}
  edge[arrin] node[lab,xshift=-1ex] {replaces} (e1);
  
\end{tikzpicture}
\end{tabular} \hspace{1cm}
&
\begin{tabular}{lllllll}
\toprule
$v_2$ & $v_4$ & $v_6$ \\
\midrule
\texttt{Michelle bachelet} & \texttt{2006-03-11} & \texttt{Ricardo Lagos} \\
\texttt{Michelle bachelet} & \texttt{2018-03-11} & \texttt{Sebastián Piñera}\\ 
\bottomrule
\end{tabular}
\end{tabular}

\caption{Graphical representation of a basic graph pattern (left), and the tabular representation of the solution mappings (right) obtained by evaluating the basic graph pattern
  over the property domain graph shown in Figure~\ref{fig:dg} \label{fig:bgp}.}
\end{figure}

\subsection{Navigational graph patterns}
A characteristic feature of graph query languages is the ability to match paths of arbitrary length that satisfy certain criteria. We call basic graph patterns enhanced with this feature \textit{navigational graph patterns}, and we define them next.

A popular way to express criteria that paths should match is through regular expressions on their labels, aka.\ \textit{regular path queries} (\textit{rpqs}). More precisely, an \textit{rpq expression} $r$ is defined by the following grammar:
\begin{eqnarray*}
  r & ::= & \varepsilon \ \mid \ o \in \objs \ \mid \  (r/r) \ \mid \ (r + r) \ \mid \ \inv{r} \ \mid \ r^*.
\end{eqnarray*}
The semantics of an rpq expression $r$ is defined in terms of its
evaluation on a property-domain graph $G$, denoted by $\semp{r}{G}$, which
returns a set of pair of nodes in the graph that are connected by
paths satisfying $r$. More precisely, assuming that $G = (O,\gamma)$,
$o \in \objs$ and $r,r_1,r_2$ are rpq expressions, we have that:
\begin{eqnarray*}
\semp{\varepsilon}{G} & = & \{(o,o) \mid o\in O\},\\
\semp{o}{G} & = & \{(o_1,o_2) \mid \exists o' \in \objs : \gamma(o') = (o_1,o,o_2)\},\\
\semp{(r_1 / r_2)}{G} & = & \{ (o_1,o_2) \mid \exists o' \in \objs : (o_1,o') \in \semp{r_1}{G} \text{ and } (o',o_2) \in \semp{r_2}{G}\}, \\
\semp{(r_1 + r_2)}{G} & = & \semp{r_1}{G}\cup \semp{r_2}{G}, \\
\semp{\inv{r}}{G} & = & \{(o_1,o_2) \mid (o_2,o_1)\in \semp{r}{G}\}.
\end{eqnarray*}
Moreover, assuming that $r^1 = r$ and $r^{n+1} = r/r^n$ for every $n \geq 1$, we have that:
\begin{eqnarray*}
  \semp{r^*}{G} & = & \semp{\varepsilon}{G} \, \cup \, \bigcup_{k \geq 1} \semp{r^k}{G}.
\end{eqnarray*}
Other rpq expressions widely used in practice can be defined by
combining the previous operators. 
In particular, $r? = \varepsilon + r$ and $r^+ = r/r^*$.


A \textit{path pattern} is a tuple $(a_1,r,a_2)$ such that $a_1, a_2 \in \objs \cup \vars$ and $r$ is an rpq expression. 
As for the case of basic graph patterns, given a path pattern $p$, we use the term $\Var{p}$ to denote the set of variables occurring in $p$. 
Moreover, the evaluation of $p = (a_1,r,a_2)$ over a property-domain graph $G$, denoted by $\semp{p}{G}$, is defined as:
\begin{eqnarray*}
  \semp{p}{G} &=& \{ \mu \mid \Dom{\mu} = \Var{p} \text{ and } (\mu(a_1),\mu(a_2)) \in \semp{r}{G} \}.
\end{eqnarray*}
For example, the expression (\texttt{Michelle Bachelet}, \texttt{(replaced by)}$^+$, $v$) is a path pattern that returns all the Presidents of Chile after Michelle Bachelet.  
Given a set $\psi$ of path patterns, $\Var{\psi}$ also denotes the set of variables occurring in $\psi$, and the evaluation of $\psi$ over a property-domain graph $G$ is defined as:
\begin{eqnarray*}
  \semp{\psi}{G} & = & \{ \mu \mid \Dom{\mu} = \Var{\psi} \text{ and } \mu_{|_{\Var{p}}} \in \semp{p}{G} \text{ for each } p \in \psi \}.
\end{eqnarray*}

A \emph{navigational graph pattern} is a triple $(V,\phi,\psi)$ where $(V', \phi)$ is a basic graph pattern for some $V' \subseteq V$, $\psi$ is a set of path patterns, and $V \subseteq \Var{\phi} \cup \Var{\psi}$. 
The semantics of a navigational graph pattern $N = (V,\phi,\psi)$ is defined as:
\begin{eqnarray*}
	\semp{N}{G} & = & \{ \mu_{|_V} \mid \Var{\mu} = \Var{\phi} \cup \Var{\psi}, \mu_{|_{\Var{\phi}}} \in \semp{\phi}{G} \ \text{ and } \ 
	\mu_{|_{\Var{\psi}}} \in \semp{\psi}{G} \}.
\end{eqnarray*}

Hence, the result of a navigational graph pattern $N = (V,\phi,\psi)$ is a set of mappings $\mu$ projected onto the set $V$ of output variables, where $\mu$ satisfies the structural restrictions imposed by $\phi$ and the path constrainst imposed by $\psi$.
Notice that multiple rpq expressions can link the same pair of nodes. This is similar to the existential semantics of path queries, as specified in the SPARQL standard \cite{HarrisS13}.

Given a domain graph $G=(O,\gamma)$, we define paths over the directed labeled graph that forms the range of $\gamma$; in other words, we do not allow for matching paths that emanate from an edge object. Such a feature could be considered in the future. We may also consider adding semantics for shortest paths, additional criteria on node or edges in the path, etc.


\subsection{Relational graph patterns}
As previously discussed (and seen in the example of Figure~\ref{fig:bgp}), graph patterns return relations (tables) as solutions. Thus we can -- and many practical graph query languages do -- use a relational-style algebra to transform and/or combine one or more sets of solution mappings into a final result.  

Towards defining this algebra, we need the following terminology.
Two mappings $\map_1$ and $\map_2$ are \emph{compatible}, denoted by $\map_1 \sim \map_2$, if $\map_1(v) = \map_2(v)$ for all variables $v$ which are in both $\Dom{\map_1}$ and $\Dom{\map_2}$. If $\map_1 \sim \map_2$, then we write $\map_1 \cup \map_2$ for the mapping obtained  by extending $\map_1$ according to $\map_2$ on all the variables in $\Dom{\mu_2}\setminus\Dom{\mu_1}$.
%
Given two sets of mappings $\Omega_1$ and $\Omega_2$, the \emph{join} 
and \emph{left outer join} between $\Omega_1$ and $\Omega_2$ are defined respectively as follows:
\begin{eqnarray*}
\Omega_1 \Join \Omega_2  &=& \{ \mu_1 \cup \mu_2 \mid \mu_1 \in \Omega_1,\, \mu_2 \in \Omega_2 \text{ and } \mu_1 \sim \mu_2 \},\\
\Omega_1 \LOJ \Omega_2  &=& (\Omega_1 \Join \Omega_2) \cup (\Omega_1 \setminus \Omega_2).
\end{eqnarray*}
With this terminology, a \textit{relational graph pattern} is recursively defined as follows:
\begin{itemize}
\item If $N$ is a navigational graph pattern, then $N$ is also relational graph pattern;
\item If $R_1$ and $R_2$ are relational graph patterns, then $(R_1 \AAND R_2)$ and $(R_1 \OPT R_2)$ are relational graph patterns.
\end{itemize}
The evaluation of a relational graph pattern $R$ over a property-domain graph $G$, denoted by $\semp{R}{G}$, is recursively defined as follows:
\begin{itemize}
\item if $R$ is a navigational graph pattern $N$, then $\semp{R}{G} = \semp{N}{G}$;
\item if $R$ is $(R_1 \AAND R_2)$ then $\semp{R}{G} = \semp{R_1}{G} \Join \semp{R_2}{G}$;
\item if $R$ is  $(R_1 \OPT R_2)$ then $\semp{R}{G} = \semp{R_1}{G} \LOJ \semp{R_2}{G}$.
\end{itemize}

\subsection{Selection conditions}
In addition to match a graph pattern against a property-domain graph, we would like to filter the solutions by imposing selection conditions over the resulting objects (i.e. nodes and edges).
More precisely, a \emph{selection condition} is defined recursively as follows: 
(a) if $v_1, v_2 \in \vars$ and $o \in \objs$, $k_1, k_2 \in \keys$, $v \in \values$ then $(v_1 = v_2)$, $(v_1 = o)$, $v_1.k_1 = v_2.k_2$ and $v_1.k_1 = v$ are selection conditions; and (b) if $C_1, C_2$ are selection conditions, then $(\neg\ C_1)$, $(C_1 \land C_2)$, $(C_1
\lor C_2)$ are selection conditions.

Given a property-domain graph $G=(O,\gamma,\lab,\prop)$, a mapping $\mu$, and a selection condition $C$, we say that $\mu$ satisfies $C$ under $G$, denoted by $\mu \models_G C$, if one of the following statements holds:
\begin{itemize}
\item
$C$ is $({v_1} = {v_2})$, $v_1, v_2 \in \Dom{\mu}$ and $\mu(v_1) = \mu(v_2)$;

\item
$C$ is $({v_1} = o_1)$, $v_1 \in \Dom{\mu}$ and $\mu(v_1) = o_1$;

\item
$C$ is $({v_1.k_1} = {v_2.k_2})$, $v_1, v_2 \in \Dom{\mu}$, 
$\prop(\mu(v_1),k_1) = \prop(\mu(v_2),k_2)$;

\item
$C$ is $({v_1.k_1} = {v})$, $v_1 \in \Dom{\mu}$ and $\prop(\mu(v_1),k_1) = v$;

\item
$C$ is $(\neg C_1)$, and it is not the case that $\mu \models_G C_1$;

\item
$C$ is $(C_1 \land C_2)$, $\mu \models_G C_1$, and $\mu \models_G C_2$;

\item
$C$ is $(C_1 \lor C_2)$ and either $\mu \models_G C_1$, $\mu \models_G C_2$, or both.
\end{itemize}

\subsection{Solution modifiers}
We consider an initial set of solution modifiers that allow for applying a final transformation on the solutions generated by a graph pattern. 
These include:
\textsf{selection}, which defines a set of elements (variables and properties) to be returned;
\textsf{order by}, which orders the solutions according to a sort criteria; and
\textsf{limit}, which returns the first $n$ mappings in a sequence of solutions.


Let $\mathbf{S}$ be the set of strings, $v \in var$ and $k \in \keys$.
A \emph{selection mapping} is a function $\tau : \mathbf{S} \to \objs \cup \values$.
A \emph{selection element} is either a variable $v$ or an expression $v.k$.   
Assume that there is a simple way to transform a selection element into a string in $\mathbf{S}$.  
Given a sequence of selection mappings $S$ and an integer $n$, the function $\texttt{limit}(S,n)$ returns the first $n$ elements of $S$ when $n > 0$, and returns $S$ otherwise.  

Given a property-domain graph $G=(O,\gamma,\lab,\prop)$, a mapping $\mu$, and a sequence of selection elements $E$, the function $\texttt{sel}(\mu,E)_G$ returns a selection mapping $\tau$ defined as follows:
if $v \in E$ then $``v" \in \dom(\tau)$ and $\tau(``v") = \mu(v)$;
if $v.k \in E$ then $``v.k" \in \dom(\tau)$ and $\tau(``v.k") = \prop(\mu(v),k)$. 
Moreover, given a set of solution mappings $\Omega$, the function $\texttt{select}$ returns a set of selection mappings defined as 
$\texttt{select}(\Omega,E)_G = \{ \texttt{sel}(\mu,E)_G \mid \mu \in \Omega \}$. 

An \emph{order modifier} is a tuple $(e,\beta)$ where $e$ is a selection element and $\beta$ is either \texttt{asc} or \texttt{desc}.
Given a sequence of selection mappings $S$ and an order modifier $o = (e,\beta)$, we say that $S$ satisfies $o$, denoted $S \models o$, if it applies that: 
(i) $\beta$ is \texttt{asc} and $S$ satisfies an ascending order with respect to $e$;
or (ii) $\beta$ is \texttt{desc} and $S$ satisfies a descending order with respect to $e$.  
Moreover, given a sequence of order modifiers $O = (o_1, \dots, o_n)$, we say that $S$ satisfies $O$, denoted $S \models O$, if it applies that:
(i)  $S \models o_1$ when $n = 1$; or
(ii) $S \models o_1$ and, for every sub-sequence of selection mappings $S' \subseteq S$ it applies that $S' \models (o_2, \dots, o_n)$ such that $\tau_i(e_1) = \tau_j(e_1)$ for any pair of selection mappings $\tau_i, \tau_j \in S'$, with $o_1 = (e_1,\beta_1)$.  

\subsection{Graph Queries}
A \textit{graph query} $Q$ is defined as a tuple ($R$, $C$, $E$, $O$, $n$), where $R$ is a relational graph pattern, $C$ is a selection condition, $E$ is a sequence of selection elements, $O = \{ o_1, \dots, o_n \}$ is a sequence of order modifiers, and $n$ is a positive integer. We assume that $R$ is the unique mandatory component. 
Given a variable $v \in \dom(R)$, the rest of components have the following expressions by default: $C$ is $v = v$, $E$ is $v$, $O$ is $(v,\texttt{asc})$ and $n = 0$. 

The evaluation of $Q$ over $G$ is defined as $\texttt{limit}(S,n)$ where   
$S = \texttt{Select}(\Omega,E)_G$,  $S \models O$, and 
$\Omega = \{ \mu_{|_V} \mid \mu \in \semp{R}{G} \land \mu \models^G C \}$.
We will assume that every graph query $Q$ = ($R$, $C$, $E$, $O$, $n$) satisfies the following two conditions:
(i)  For every sub-pattern $R' = (R_1 \OPT R_2)$ of $R$ and for every variable $v$ occurring in $R$, it applies that, if $v$ occurs both inside $R_2$ and outside $R'$, then it also occurs in $R_1$;
(ii) It applies that $\var(C) \subseteq \var(R)$.
Then, we say that $Q$ is a \emph{well-designed graph query}.

We finish this section noting that the semantics of a declarative query expression: 

\begin{dql}
	SELECT E
	MATCH  R
	WHERE  C
	ORDER BY O 
	LIMIT n
\end{dql}

\noindent is defined as the outputs of the graph query 
($R$, $C$, $E$, $O$, $n$).

\end{document}